\pdfoutput=1
\documentclass[global,twocolumn]{svjour}
\usepackage{graphicx}% Include figure files
\usepackage{dcolumn}% Align table columns on decimal point
\usepackage{bm}% bold math
\usepackage{amsmath}
\usepackage{hhline}
\usepackage{footnote}
\newcommand{\HD}{HD$^+$}
\newcommand{\Be}{Be$^+$}
\newcommand{\vv}{(0,2)$\rightarrow$(8,3)}
\newcommand{\vvL}{(\emph{v,L}):(0,2)$\rightarrow$(8,3)}
\newcommand{\HHD}{H$_2$D$^+$}
\newcommand{\HDD}{HD$_2^+$}
\newcommand{\HHH}{H$_3^+$}

\newcommand{\Tz}{\bar{T}_0}
\newcommand{\Tzz}{$\bar{T}_0$}

\newcommand{\ket}[1]{%
     | #1 \rangle}

\newcommand{\epratio}[1]{%
     $m_p/m_e\:$}
\newcommand{\dpol}[1]{%
     \alpha_{vJM}(\omega)}
\newcommand{\dpolAv}[1]{%
     \overline {\alpha_{vJ}(\omega)}}
\newcommand{\dpolO}[1]{%
     \alpha_{vJM}(0)}
\newcommand{\dpole}[1]{%
     \alpha^e_{vJM}(\omega)}
\newcommand{\dpoleAv}[1]{%
     \overline{\alpha^e_{vJ}(\omega)}}
\newcommand{\dpoleO}[1]{%
     \alpha^e_{vJM}(0)}
\newcommand{\dpolrv}[1]{%
     \alpha^{rv}_{vJM}(\omega)}
\newcommand{\dpolrvAv}[1]{%
     \overline{\alpha^{rv}_{vJ}(\omega)}}
\newcommand{\dpolrvO}[1]{%
     \alpha^{rv}_{vJM}(0)}

\DeclareGraphicsExtensions{.pdf,.png,.jpg}

\begin{document}

%%\preprint{APS/123-QED}

\title{High-precision spectroscopy of the HD$^+$ molecule at the 1-p.p.b. level}% Force line breaks with \\

\author{J. Biesheuvel\inst{1}\thanks{\emph{Present address: Statistics Netherlands (CBS), P.O. Box 24500, 2490 HA The Hague, The Netherlands
}}, J.-Ph. Karr\inst{2,3}, L. Hilico\inst{2,3}, K.S.E. Eikema\inst{1}, W. Ubachs\inst{1}, J.C.J. Koelemeij\inst{1}}
\institute{LaserLaB, Department of Physics and Astronomy, Vrije Universiteit, De Boelelaan 1081, 1081 HV Amsterdam, The Netherlands\\ \and
Laboratoire Kastler Brossel, UPMC-Sorbonne Universit\'es, CNRS, ENS-PSL Research University, Coll\`ege de France, 4 place Jussieu, F-75005 Paris, France \and Departement de physique, Universit\'e d'Evry Val d'Essonne, Boulevard Fran\c cois Mitterrand, F-91025 Evry cedex, France}
\offprints{j.c.j.koelemeij@vu.nl}          % Insert a name or remove this line

\date{Received: \today / Revised version: date}%
\titlerunning{High-precision spectroscopy of the HD$^+$ molecule at the 1-p.p.b. level}%\ldots}
\authorrunning{J. Biesheuvel \textit{et al.}}
\maketitle
\begin{abstract}
Recently we reported a high precision optical frequency measurement of the $(v,L)$: (0,2)$\rightarrow$(8,3) vibrational overtone transition in trapped deuterated molecular hydrogen (\HD ) ions at 10~mK temperature. Achieving a resolution of 0.85 parts-per-billion (p.p.b.) we found the experimental value ($\nu_0= 383,407,177.38(41)$~MHz) to be in agreement with the value from molecular theory ($\nu_{\text{th}}=383,407,177.150(15)$~MHz) within 0.6(1.1) p.p.b. [Biesheuvel \textit{et al.}, Nat. Commun. \textbf{7}, 10385 (2016)]. This enabled an improved test of molecular theory (including QED), new constraints on the size of possible effects due to 'new physics', and the first determination of the proton-electron mass ratio from a molecule. Here, we provide the details of the experimental procedure, spectral analysis, and the assessment of systematic frequency shifts. Our analysis focuses in particular on deviations of the \HD\ velocity distribution from thermal (Gaussian) distributions under the influence of collisions with fast ions produced during (laser-induced) chemical reactions, as such deviations turn out to significantly shift the hyperfine-less vibrational frequency as inferred from the saturated and Doppler-broadened spectrum, which contains partly unresolved hyperfine structure. %In particular we find that deviations from a thermal (Gaussian) distribution, due to laser-induced chemistry, leads to a lineshift. We perform realistic molecular dynamics simulations to obtain an estimate of the relative frequency correction, which is found to be -0.6(6) p.p.b.. %The agreement between our measurement and theory  is not only the most stringent test of higher order QED corrections molecular theory (including high-order QED), but also improves constraints on hypothetical fifth forces and compactified dimensions. Furthermore, this work enables the first determination of the proton-to-electron mass ratio from a molecular system.

%\begin{description}
%\end{description}
\end{abstract}

%%\pacs{Valid PACS appear here}% PACS, the Physics and Astronomy
\maketitle

%\tableofcontents

\section{\label{intro}Introduction}
Because of their relative simplicity, molecular hydrogen ions such as H$_2^+$ and HD$^+$ are benchmark systems for molecular theory~\cite{Korobov2014a,Korobov2014b}, and suitable probes of fundamental physical models~\cite{Salumbides2013,Salumbides2015}. Already four decades ago, Wing \emph{et al}.~\cite{Wing1976} performed measurements of rovibrational transitions in \HD\ and conjectured that such measurements could be used to test quantum electrodynamics (QED) theory in molecules, and lead to an improved value of the proton-to-electron mass ratio, $\mu$. Today, QED corrections to ro-vibrational transition frequencies in H$_2^+$ and HD$^+$ have been calculated up to the order $m_e \alpha^8$, with $m_e$ the electron mass and $\alpha$ the fine-structure constant, leading to relative uncertainties of the order of $3$--$4\times 10^{-11}$ ~\cite{Korobov2014a,Korobov2014b}. The frequencies of these transitions are typically in the (near-) infrared with linewidths below 10~Hz, which makes them amenable to high-resolution laser spectroscopy. Optical clocks based on trapped atomic ions have shown that laser spectroscopy at very high accuracy (below one part in $10^{17}$) is possible~\cite{Chou2010}. Recent theoretical studies point out that also for molecular hydrogen ions experimental uncertainties in the $10^{-16}$ range should be possible ~\cite{Schiller2014,Karr2014}.
%High-precision measurements on molecular hydrogen ions can also be used to probe physics beyond the standard model, which could manifest itself in the form of hypothetical fifth forces between hadrons~\%cite%{Salumbides2013,%Salumbides2014} or gravitational effects from compactified higher dimensions~\cite{Salumbides2015}. If present, such effects might result in a discrepancy between molecular theory and experiment.

Recent frequency measurements of rovibrational transitions in \HD\ were done with a reported uncertainty of 1--2~p.p.b.~\cite{Koelemeij2007a,Bressel2012}. However, the most precise measurement of a rovibrational transition in HD$^+$ so far resulted in a 2.7~p.p.b. (2.4$\sigma$) discrepancy with more accurate theoretical data \cite{Bressel2012}. The question whether this offset is a statistical outlier, or caused by an experimental systematic effect or possible 'new physics' has remained unanswered.  Therefore, additional experimental data on HD$^+$  are needed.  In this article we present the result of a high-precision frequency measurement of the rovibrational transition \vvL\ in the HD$^+$  molecule, and compare it with state-of-the-art molecular theory to find excellent agreement. From this comparison we subsequently derive new constraints on the size of effects due to possible new physics in HD$^+$, and we determine a value of $\mu$ for the first time from a molecular system~\cite{Biesheuvel2015}.

This article is organized as follows. In Sec.~\ref{theory} we briefly review the theory of HD$^+$ relevant to this experiment. In Sec.~\ref{expsetup} we describe a setup for spectroscopy of trapped HD$^+$ ions, sympathetically cooled by laser-cooled beryllium ions.  The analysis of the spectroscopic data, systematic frequency shifts and the results are discussed in Sec.~\ref{results}, followed by the conclusions in Sec.~\ref{conclusion}.

\section{\label{theory}Theory}
\subsection{Calculation of ro-vibrational frequency transitions in \HD}\label{hfless}
The calculation of ro-vibrational energies in quantum mechanical three-body systems is usually divided into a non-relativistic part, complemented by the relativistic and radiative parts, and finite nuclear size contributions. The non-relativistic part is calculated through solving the three-body Schr\"o\-ding\-er equation, which can be done with practically infinite precision~\cite{Korobov2000} (up to a relative precision of $\sim10^{-32}$ ~\cite{Li2007,Hijikata2009}). The resulting wavefunctions allow an analytical evaluation of the Breit-Pauli Hamiltonian and the leading-order radiative corrections.

Recently, the precision of relativistic and radiative energy corrections to the non-relativistic energies was strongly improved. With the inclusion of the full set of contributions of order $m \alpha^7$ and leading-order terms of order $m \alpha^8$, the relative uncertainty is now below $4\times 10^{-11}$. For example, the theoretically determined value of the \HD\ \vvL\ transition frequency, $\nu_{\text{th}}$, is  383,407,177.150(15)~MHz, and has a relative uncertainty of $4 \times 10^{-11}~$\cite{Korobov2014a,Korobov2014b}. Note that the specified error (within parentheses) does not include the uncertainty of the fundamental constants used. By far the largest contribution is due to the $4.1 \times 10^{-10}$ uncertainty of the 2010 Committee on Data for Science and Technology (CODATA) value of $\mu$~\cite{CODATA2010}, which translates to a frequency uncertainty of 59~kHz.

\subsection{Hyperfine structure and rotational states}\label{hfstr}%(gebruik evt. fig1abc van Jeroen, voor de hele sectie Theory)
Since the \HD\ constituents possess nonzero spin, the ro-vibrational transition spectra contain hyperfine structure due to spin-spin and spin-orbit couplings. In \cite{Bakalov2006} the hyperfine energy levels and eigenstates in \HD\ are calculated by diagonalization of the effective spin Hamiltonian
\begin{eqnarray}\label{eq:hyperfine1}
H_{\text{eff}} & = &E_1(\mathbf{L} \cdot \mathbf{s_e}) + E_2(\mathbf{L} \cdot \mathbf{I_p}) + E_3(\mathbf{L} \cdot \mathbf{I_d})+E_4(\mathbf{I_p} \cdot \mathbf{s_e})
\nonumber \\
&& +\:  E_5(\mathbf{I_d} \cdot \mathbf{s_e}) + E_6 K_d(\mathbf{L},\mathbf{I_p},\mathbf{s_e})+  E_7 K_d(\mathbf{L},\mathbf{I_d},\mathbf{s_e}) \nonumber\\
&& +\:E_8 K_d(\mathbf{L},\mathbf{I_p},\mathbf{I_d})+ E_9 K_Q(\mathbf{L},\mathbf{I_d}),
\end{eqnarray}
where the spin coefficients, $E_i$, are obtained by averaging the Breit-Pauli Hamiltonian over the nonrelativistic wavefunctions, $\mathbf{L}$ is the rotational angular momentum operator, and  $\mathbf{s_e}$, $\mathbf{I_p}$ and $\mathbf{I_d}$ are the electron, proton and deuteron spin operators. $K_d$ and $K_Q$ are spherical tensors composed of angular momenta, whose explicit form is given in \cite{Bakalov2006}. The strongest coupling is the proton-electron spin-spin interaction (the term in $E_4$ in Eq.~(\ref{eq:hyperfine1})), and the preferred angular momentum coupling scheme is
\begin{equation}\label{eq:hyperfine2}
\mathbf{F}= \mathbf{I_p}+\mathbf{s_e} \qquad \mathbf{S}= \mathbf{F}+\mathbf{I_d} \qquad \mathbf{J}= \mathbf{L}+\mathbf{S}.
\end{equation}
For states with $L=0$, $L=1$ and $L\geq 2$ this leads to the presence of 4, 10 and 12 hyperfine levels, respectively, which are distributed over an energy range of $\sim$1~GHz (Fig.~\ref{REMPDabc}(b)). %, as schematically depicted in Fig.~\ref{picHyperfine}.%%%%Dit nog aanpassen, samenvoegen met andere figuur?? JK%%%%%
Diagonalization produces eigenstates $\phi_{\tilde{F}\tilde{S}JM_J}$, with the magnetic quantum number $M_J$ corresponding to the projection of $J$ onto the laboratory-fixed z-axis. Note that after diagonalization the quantum numbers $\tilde{F}, \tilde{S}$ are only approximately good, and a hyperfine eigenstate can be expressed in the 'pure' basis states $\ket{FSJM_J}$ in the form
\begin{equation}\label{HFeigenstates}
\phi_{\tilde{F}\tilde{S}JM_J}=\sum_{F,S}{\beta_{FSJ}\ket{FSJM_J}},
\end{equation}
with real-valued coefficients $\beta_{FSJ}$. In \cite{Bakalov2006} the hyperfine levels in vibrationally excited states in \HD\ are calculated with an error margin of $\sim$~50~kHz. This uncertainty might propagate through a spectral analysis employing a lineshape model which includes the theoretical hyperfine structure (as we do below). However, the uncertainty contribution to the determined hyperfine-less \vvL\ transition frequency is expected to be at least five times smaller due to strong correlation and partial cancelation of the error (see also Sec.~\ref{nu0results}).
%\begin{figure}
%\centering
%\includegraphics[width=10cm]{picHyperfine}% Here is how to import EPS art
%\caption{\label{picHyperfine}Schematic diagram of hyperfine states of the lower vibrational energy levels in %\HD\ with $L \ge 2$. }
%\end{figure}

\subsection{Determination of transition rates}\label{rates}
Because of the hyperfine structure, the spectrum of the \vv\ electric dipole transition consists of a large number of hyperfine components. Together with Doppler broadening, this leads to an irregular lineshape. In addition, the excitation laser may address multiple hyperfine states simultaneously, which are furthermore coupled to other rotational states by the ambient 300~K blackbody radiation (BBR) field. Therefore, for the analysis of the \vv\ signal we develop a model based on Einstein rate equations which takes all resonant molecule-electric field interactions into account. We note that at 300~K, states with $v=0$ and $L=1$--6 are significantly populated, with 27\% in $L=2$, and 2.4\% in states with $L\geq 6$. %(see Fig.~\ref{LBBRdistr}).
%\begin{figure}
%\centering
%\includegraphics[width=6cm]{LBBRdistr}% Here is how to import EPS art
%\caption{\label{LBBRdistr} (Color online) Bar chart of the $L$ occupancy of \HD\ at 300~K for $L=0$ to $L=6$. %Higher $L$ states obviously are less occupied then $L=6$ and are not shown here.}
%\end{figure}
Below, we calculate the Einstein rate coefficients at the level of individual hyperfine states for transitions driven by the laser and BBR fields.

Following the approach of Koelemeij~\cite{Koelemeij2012}, we first ignore hyperfine structure and %solve the radial Schr\"odinger equation which gives the radial wave function of nuclear motion $\chi_{v,L}$:
%\begin{eqnarray}\label{eq:radSchreq}
%-\frac{\hbar^2}{2 \mu}\frac{d^2}{dR^2}\chi_{v,L}(R)+\left[V(R)+\frac{\hbar^2L(L+1)}{2\mu_{\text{red}} R^2} \right]\chi_{v,L}(R)\nonumber \\
% =  E_{v,L}\chi_{v,L}(R)
%\end{eqnarray}
%where $R$ is the nuclear separation, $\mu_{\text{red}}$ stands for the reduced mass of the molecule, $v$ labels vibrational state, $L$ is the angular momentum and $E_{vL}$ is the ro-vibrational energy. $V(R)$ is the potential energy curve belonging to the 1s$\sigma$ electronic ground state of \HD\ which is shown in Fig.~\ref{REMPDabc}a. The transition dipole moment is then given by
obtain transition dipole moments given by
\begin{equation}
\mu_{v v^{\prime}L L^{\prime}}= \int_0^{\infty} \chi_{v',L'}(R)D_1(R)\chi_{v,L}(R)dR ,
\end{equation}
where $\chi_{v,L}$ denotes the radial wave function of nuclear motion obtained by numerical solution of the radial Schr\"o\-ding\-er equation, $R$ stands for the internuclear separation, and $D_1(R)$ denotes the permanent electric dipole moment function of the $1s\sigma$ electronic ground state. %which is calculated along with the potential energy curve in~\cite{Esry1999}.

To take hyperfine structure into account we evaluate the dipole transition matrix element between two states $\phi_{\tilde{F}\tilde{S}JM_J}\chi_{v,L}$ and $\phi_{\tilde{F'}\tilde{S'}J'M'_J}\chi_{v',L'}$   \cite{Brown2003,Koelemeij2012,Koelemeij2011}
\begin{eqnarray} \label{eq:hflinestrength}
\lefteqn{\langle\phi_{\tilde{F}\tilde{S}JM_J}\chi_{v,L}| \vec{E} \cdot \vec{\mu} |\phi_{\tilde{F'}\tilde{S'}J'M'_J}\chi_{v',L'}\rangle} \nonumber \\
&=& \sum_{F'S'} \beta_{FSJ}\beta^*_{F'S'J'} E_p (-1)^{J+J'+S-M_J+1+L'+L} \nonumber \\
&& \times\: [(2J+1)(2J'+1)(2L+1)(2L'+1)]^{1/2} \nonumber \\
&& \times \left( \begin{array}{ccc}
L & 1 & L' \\
M_J & p & M_J-p
\end{array} \right)
\left( \begin{array}{ccc}
L & 1 & L' \\
0 & 0 & 0
\end{array} \right)
\left\{ \begin{array}{ccc}
L' & J' & S \\
J & L & 1
\end{array} \right\}\nonumber \\
&& \times\: \mu_{vv'LL'}.
\end{eqnarray}
Equation~(\ref{eq:hflinestrength}) includes a transformation from the molecule-fixed frame to the laboratory-fixed frame~\cite{Brown2003}, with $p$ denoting the polarization state of the electric field with respect to the laboratory-fixed frame ($p\in \{-1,0,1\}$).

The linestrength is subsequently calculated by squaring and summing over the $M_J$ states \cite{Hilborn1982}:
\begin{equation}\label{eq:linestrength1}
S_{\alpha \alpha'}= \sum_{\substack{M_J}} |\langle\phi_{\tilde{F}\tilde{S}JM_J}\chi_{v,L}| \vec{E} \cdot \vec{\mu} |\phi_{\tilde{F'}\tilde{S'}J'M'_J}\chi_{v',L'}\rangle|^2,
\end{equation}
with $\alpha \equiv  vL\tilde{F}\tilde{S}J$.
The Einstein rate coefficients for spontaneous emission, absorption and stimulated emission are subsequently obtained as described in \cite{Tran2013}, and used in Sec.~\ref{lsmodel} to calculate the expected \vv\ spectrum.
%($A_{\alpha \alpha'}$, $\bar{B}_{\alpha' \alpha}$ and $B_{\alpha \alpha'}$) are obtained through the relations
%\begin{equation}\label{eq:einsteincoeff1}
%A_{\alpha \alpha'}= \frac{2 \omega^3_{vv'LL'}}{3 \epsilon h c^3}\frac{S_{\alpha \alpha'}}{2J'+1}
%\end{equation}
%\begin{equation}\label{eq:einsteincoeff2}
%B_{\alpha \alpha'}= \frac{\pi^2 c^3 }{\hbar \omega^3_{vv'LL'}}A_{\alpha \alpha'}
%\end{equation}
%\begin{equation}\label{eq:einsteincoeff2}
%\bar{B}_{\alpha' \alpha}= \frac{2J'+1}{2J+1}B_{\alpha \alpha'}.
%\end{equation}
%Note that we use the hyperfineless transition frequency $\omega_{vv'LL'}$ rather than $\omega_{\alpha \alpha'}$, which has a negligible effect on the value of the rate coefficients.
%These results are used in Sec.~\ref{lsmodel} to calculate the expected \vv\ signal, which is incorporated into a model describing the dynamics during the spectroscopic measurement.

\section{\label{expsetup}Experiment}
\subsection{Trapping and cooling Be$^+$ and HD$^+$}
\begin{figure}
\centering
\includegraphics[width=10cm]{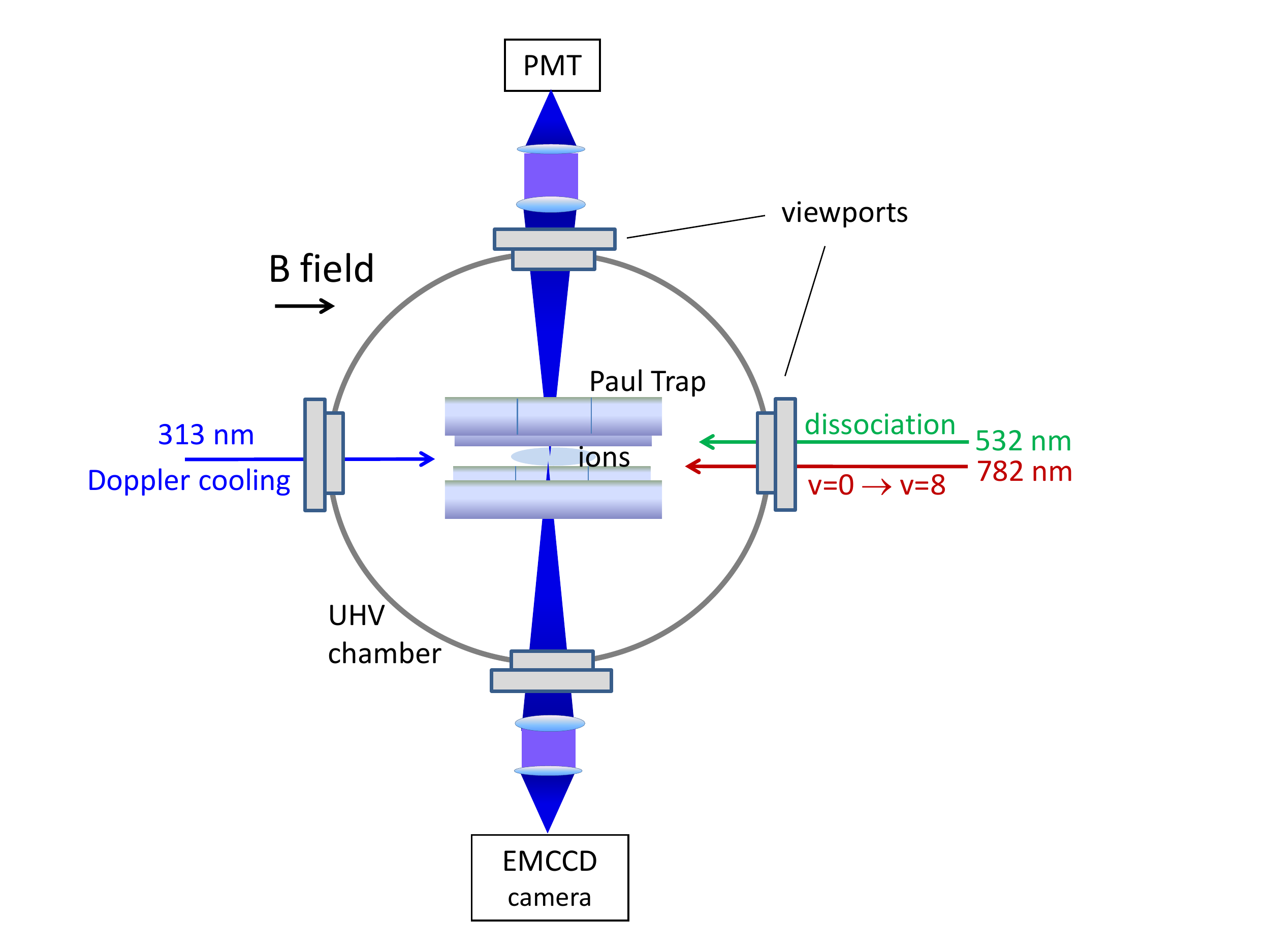}% Here is how to import EPS art
\caption{\label{setup2} (Color online) Schematic view of the trap setup. An ultrahigh vacuum (UHV) chamber houses a linear rf trap in which \Be\ and \HD\ ions are loaded by electron-impact ionization. \Be\ ions are Doppler-cooled by the 313 nm laser, which is directed parallel to the trap axis and $B$-field direction. The 313 nm \Be\ fluorescence is imaged onto and detected with a PMT and an EMCCD camera. The 782 nm and 532 nm cw REMPD lasers are overlapped with the 313 nm laser and the ions, but propagate in the opposite direction.}
\end{figure}

To achieve narrow linewidths and small systematic shifts, we choose to perform spectroscopy on small samples of \HD\ molecules in a radiofrequency (rf) ion trap. We reduce the motional temperature of the \HD\ ions by storing them together with \Be\ ions which are Doppler-cooled by a continuous-wave (cw) 313 nm laser beam (see \cite{Koelemeij2012,Biesheuvel2013,Cozijn2013} for details). The rf trap is placed inside an ultra-high vacuum chamber with a background pressure of $1\times10^{-10}$~mbar. The rf trap operates at a frequency of 13.2~MHz, leading to Be$^+$ radial trap frequencies of $\omega_r=2\pi \times 290$~kHz. The trap geometry and rf circuitry are designed so as to minimize unwanted rf fields and phase differences between the rf electrodes. The two dc electrodes are segmented into two endcaps and a center electrode (Fig. \ref{setup2}).
The dc voltages of the center electrodes, rf electrodes and endcap pairs can be individually adjusted to compensate stray electric fields. \Be\ and \HD\ are loaded by electron-impact ionization as done by Blythe \textit{et al.} \cite{Blythe2005}, and monitored with a photomultiplier tube (PMT) and an electron-multiplied charge-coupled-device (EMCCD) camera. EMCCD images show ellipsoidal mixed-species Coulomb crystals, with a dark core of molecular hydrogen ions surrounded by several shells of fluorescing \Be\ ions (Fig.~\ref{SimEMCCDions}). The apparatus and procedures for loading and compensation of stray electric fields are described in more detail in~\cite{Koelemeij2012}.

\subsection{Spectroscopy of HD$^+$}\label{spectroscopy}
\begin{figure*}
\centering
\includegraphics[width=14cm]{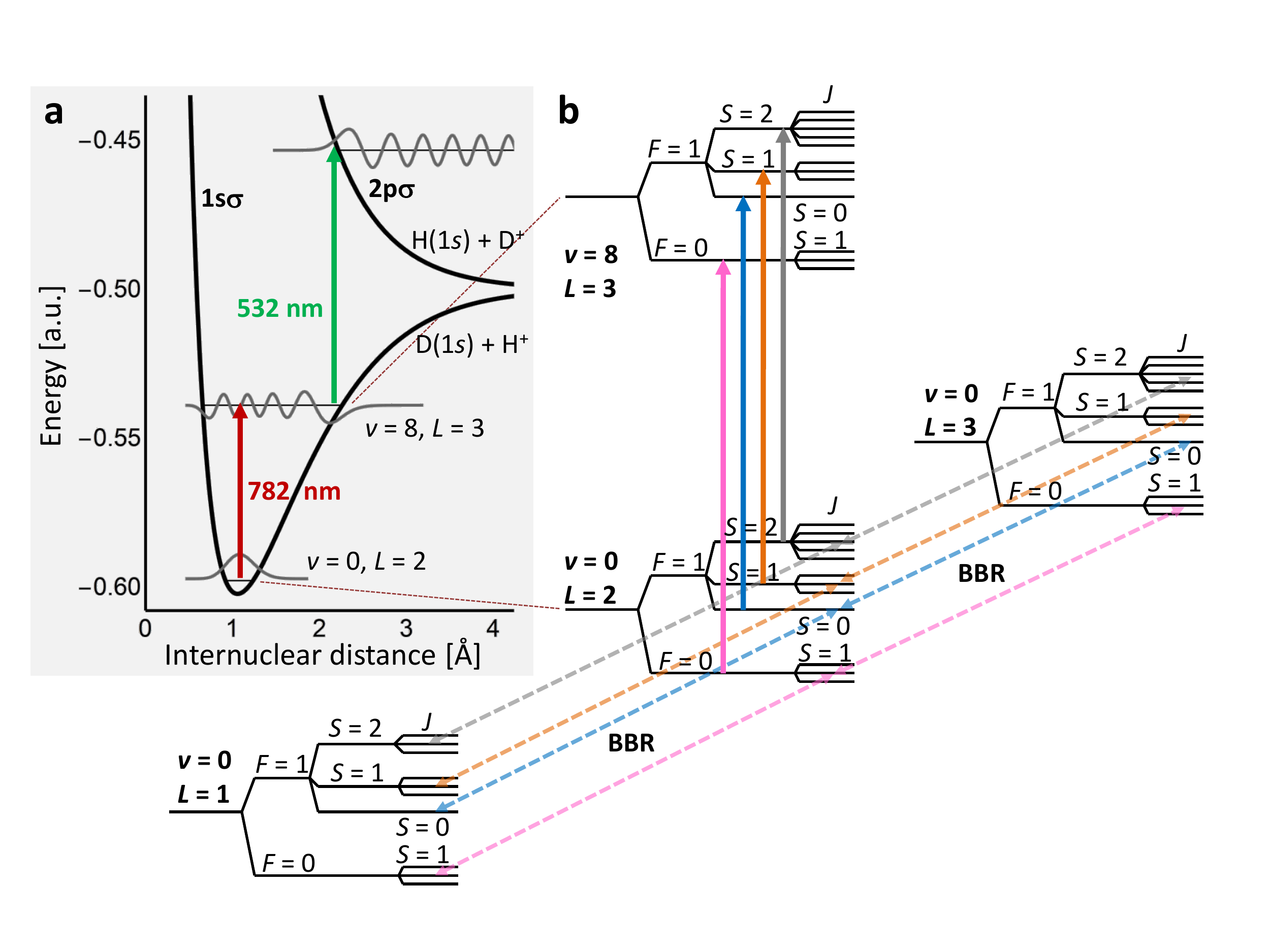}% Here is how to import EPS art
\caption{\label{REMPDabc}(a) (Color online) Potential energy curves of the 1s$\sigma$ and 2p$\sigma$ electronic states are plotted together with the radial nuclear wave functions for $v$=0, $v$=8, and for the dissociative wave function in the 2p$\sigma$ state.The REMPD scheme is also shown, with the red arrow indicating the \vvL\ transition by the 782 nm laser and the green arrow corresponding to the 532 nm photon which dissociates the molecule. (b) Detailed partial level scheme (including hyperfine levels) showing the \vv\ transition together with the BBR interaction which couples the rotational levels $L$= 1,2,3 through electric-dipole transitions. Coupling to $L=0$ and $L=4,5,6$ also occurs, but is not shown here.}
\end{figure*}

The \vv\ transition in \HD\ is detected destructively through resonance enhanced multi-photon dissociation (REMPD), see Fig. \ref{REMPDabc}a.
A 782~nm cw titanium:sap\-phire laser is used to excite \HD\ from its vibrational ground state to the \emph{v}=8 state, which is efficiently dissociated by the field of a co-propagating 532~nm cw laser beam. Both lasers are directed along the trap axis and counter-propagate the 313~nm laser (see Fig.~\ref{setup2}). Since all HD$^+$ ions are initially in the vibrational ground state (which is too deeply bound to be dissociated by the 532~nm laser), dissociation only takes place if the 782~nm laser is resonant with the \vv\ transition. We used 90~mW of 532~nm radiation, focused to a beam waist of 140~$\mu$m, which is sufficient to dissociate an HD$^+$ molecule from this level within a few ms, much faster than the spontaneous decay of the $v$=8 state (lifetime 12~ms). During each measurement cycle, the \HD\  ions are exposed to the REMPD lasers for 10~s. During the first seconds, the majority of the HD$^+$ in $L$=2 are dissociated, leading to depletion of the $L$=2 state (Fig.~\ref{REMPDabc}). During the remainder of the REMPD period, BBR repopulates the $L$=2 state from other rotational levels, which enhances the number of dissociated \HD\ ions by approximately a factor of two.

The 782~nm laser has a linewidth of $\sim$0.5 MHz and is frequency-stabilized by locking its frequency to a nearby mode of a self-referenced optical frequency comb laser. To this end, a 63.5~MHz beat note is created by mixing the light of the 782 nm laser and the frequency comb. The frequency comb itself is locked to a rubidium atomic clock for short-term stability, which is disciplined to the 1-pps signal of a GPS receiver for long-term accuracy and traceability to the SI second with 2$\times$10$^{-12}$ relative uncertainty.

\begin{figure*}
\centering
\includegraphics[trim=0cm 5cm 0cm 6cm, clip, width=17cm, angle=0]{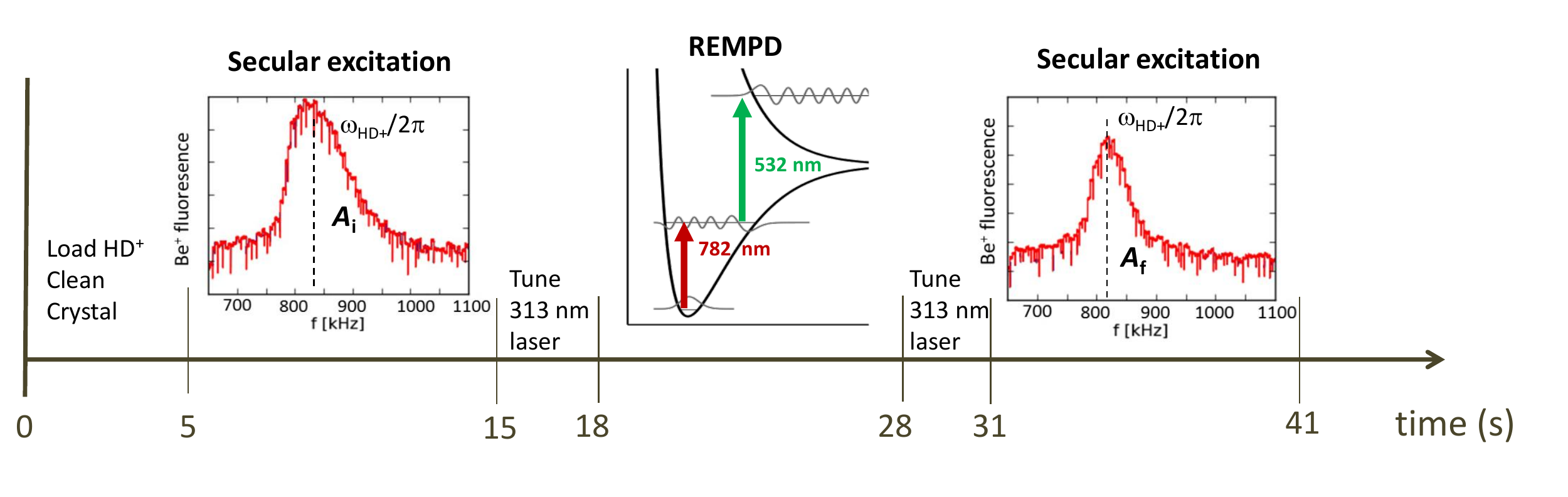}% Here is how to import EPS art
\caption{\label{fig:ssREMPDss} The secular scan - REMPD - secular scan detection scheme. At the start of each cycle a new sample of \HD\ ions is loaded into the ion trap. During loading the \Be\ ions are exposed to neutral HD gas, leading to the formation of a small number of BeH$^+$ and BeD$^+$ ions (see also Table~\ref{tab:chemreactions}). These are expunged from the trap by applying a dc quadrupole potential of $\sim 0.9$~V, which reduces the trap depth such that only ions with mass $\leq 9$~amu remain trapped. Five seconds after loading, the \HD\ is motionally excited by scanning an rf electric field over the secular motional resonance frequency at 830~kHz. A 313~nm laser detuning of $-300$~MHz is used, in which case the secular excitation results in a rise in the \Be\ fluorescence. Ten seconds later, the 313~nm laser is tuned to $-$80~MHz from the \Be\ cooling transition, and its intensity reduced to $I\sim 2I_{\text{sat}}$. After 10~s of REMPD, the 313~nm laser settings are restored to their values as used for the first secular scan, and a second secular scan is executed.  A smaller 313~nm fluorescence peak indicates loss of \HD.}
\end{figure*}

During REMPD, about 300~mW of 782~nm light is used with a beam waist of $\sim$120~$\mu$m at the location of the ions. The 313~nm cooling laser is detuned by $-80$~MHz from the cooling resonance and reduced in power to $\sim$70~$\mu$W, which results into an intensity of two times the saturation intensity of the 313~nm cooling transition, and leads to an ion temperature of about 10~mK. To obtain a measure of the fraction of \HD\ ions lost due to REMPD, we employ so-called secular excitation~\cite{Roth2006}. This procedure is based on the indirect heating of the \Be\ which occurs when the motion of the \HD\ ions is resonantly excited by an additional radial rf field as it is scanned over the resonance frequency. The heating of \Be\ leads to a change of the 313~nm fluorescence level, which is connected to the number of \HD\ ions.  A single measurement cycle consists of secular scan (10~s), followed by 10~s of REMPD and another 10-s secular scan (see Fig.~\ref{fig:ssREMPDss}) to infer the number of \HD\ ions remaining after REMPD. To obtain a spectrum, this procedure is repeated for different frequencies of the 782~nm laser (indicated by the variable $\nu$), which are chosen as follows. The $\sim$25 strongest hyperfine components of the $(0,2)\rightarrow (8,3)$ transition are located in the range ($-110$~MHz, 140~MHz) around the hyperfine-less frequency. As we expect Doppler broadening to $\sim$16~MHz, we divide this range into a set of 140 evenly spaced frequencies (spacing $\sim 1.8$~MHz) at which REMPD spectroscopy is performed. To convert possible time-varying systematic effects into random noise, we randomize the ordering of the frequency list. For each frequency point, six to seven REMPD measurements are made, which results into a spectrum consisting of 886 REMPD measurements in total.

Scanning the frequency of the radial rf field over the secular motional resonance of HD$^+$ ($\sim$830 kHz) temporarily heats up the Coulomb crystal to a few Kelvin. At a detuning of the 313~nm cooling laser of $-300$~MHz, and using a few mW of laser power (corresponding to $\sim 80$ times the saturation intensity), such a 'secular scan' shows up as a peak in the 313~nm beryllium fluorescence versus rf frequency (Fig.~\ref{fig:ssREMPDss}). In Appendix~\ref{nlfunction} we show that the area under this peak, $A$, scales with the number of HD$^+$ ions, although not linearly. We define a spectroscopic signal as the relative difference between the areas of the initial and final secular scan peaks:
\begin{equation}\label{S}
S=\frac{A_i-A_f}{A_i}
\end{equation}
Repeating this procedure for all 140 pre-selected frequencies of the 782~nm laser while recording fluorescence traces with both the PMT and EMCCD camera, we obtain a spectrum consisting of 1772 data points $(\nu_j, S_j)$ (with $j=1 \ldots 1772)$.

\section{\label{results}Results and Discussion}
\subsection{\label{lsmodel}Spectral lineshape model}
In order to determine the hyperfine-less rovibratiotional \vv\ transition frequency, $\nu_0$, we need a realistic lineshape model which includes the relevant physics present during REMPD. Here the aim is to obtain a spectral lineshape function in which all effects are parameterized. Parameter values are estimated by independent means where possible, or included as a fit parameter otherwise. Importantly, the fit function will contain a variable $\nu-\nu_{0,\text{fit}}$, where $\nu_{0,\text{fit}}$ is a fit parameter from which we later deduce the value of $\nu_0$. Before fitting, the $(\nu,S)$ data are corrected for reactions with background gas (primariliy H$_2$). This procedure is described in Sec.~\ref{bg}.

We start with building a state vector $\boldsymbol{\rho}(t)$ which contains the population in all 62 hyperfine levels in the rotational states ranging from  $L$=0 to $L$=5 with $v$=0. This includes 97.6\% of the total internal states of $v$=0 given a BBR temperature of 300~K. We neglect the Zeeman splitting due to 0.19~mT B-field at the location of the trapped ions, as this splitting is negligibly small compared to the Doppler linewidth and the width of the BBR spectrum. The lineshift to $\nu_0$ due to the Zeeman effect is considered separately in Sec.~\ref{Zeeman}. Also the Stark effect, the electric-quadrupole shift and second-order Doppler shift are not included in this model, but addressed in Sec.~\ref{Zeeman}.

During the 10~s of REMPD the hyperfine levels in the $L$=2 initial state interact with the 782 nm laser and BBR. We here make the simplifying assumption that any population in the $v=8$ target state will be dissociated by the 532~nm laser. The interaction with the 782 nm laser is therefore modeled as a simple loss process. The evolution of the state vector $\boldsymbol{\rho}(t)$ is obtained by solving the set of coupled rate equations:
\begin{equation}\label{eq:rateeq}
\frac{d \boldsymbol{\rho}(t)}{d t}=M_{\text{R}}\ \boldsymbol{\rho}(t) +M_{\text{BBR}} \ \boldsymbol{\rho}(t),
\end{equation}
where $M_{\text{R}}$ and $M_{\text{BBR}}$ are the matrices describing the interaction with the REMPD lasers and the BBR field. Generally, the diagonal elements of these matrices contain depopulating (negative) terms, whereas the off-dia\-go\-nal elements describe the population a certain state gains from another. Since $M_{\text{R}}$ describes losses it only contains diagonal entries of the form
\begin{equation}\label{eqMrempd}
M_{\text{R},\alpha \alpha} = -\sum_{\alpha'} B_{\alpha \alpha'}D_z \left((\omega-\omega_{\alpha\alpha'}),T_{\text{HD}^+} \right) I_{\text{laser}}/c,
\end{equation}
where $B_{\alpha \alpha'}$ denotes the Einstein coefficient for absorption from a lower hyperfine-rovibrational level $\alpha$ to an upper level $\alpha'$ (which for $M_\text{R}$ the states $\alpha$ and $\alpha'$ being restricted to those having $v=0,L=2$ and $v'=8,L'=3$, respectively). The corresponding transition frequency is $\omega_{\alpha \alpha'}$, $I_{\text{laser}}$ is the intensity of the 782 nm laser, and $D_z$ represents a normalized response function averaged over the distribution of $z$-velocities of the \HD\ ions. This involves an integration over all Doppler shifts $-k_z v_z$ observed by the \HD\ ensemble, where $k_z$ is the wavevector of the 782~nm laser and $v_z$ the velocity in the $z$-direction. If the \HD\ velocity distribution is thermal (Gaussian), $D_z$ depends only on the temperature in the $z$-direction, $T_{\text{HD}^+}$. However, the effects of micromotion and chemistry in the Coulomb crystal during REMPD lead to deviations from a thermal distribution. This implies that $D_z$ cannot be described by a single Gaussian lineshape. In Secs.~\ref{mm} and \ref{chemistry} these processes are explained in detail and the precise shapes of $D_z$ are determined.

The matrix $M_{\text{BBR}}$ contains both diagonal and off-diagonal elements, taking into account the rate of exchange of population between all involved levels $\alpha$ and $\alpha'$ mediated through all possible electric-dipole transitions,
\begin{equation*}
\begin{split}
&A_{\alpha' \alpha}, \\
&B_{\alpha' \alpha}W(\omega_{\alpha \alpha'},T_{\text{BBR}}), \\
&B_{\alpha \alpha'}W(\omega_{\alpha \alpha'},T_{\text{BBR}}),
\end{split}
\end{equation*}
which correspond to spontaneous emission, stimulated emission and absorption by BBR, respectively. The relationship with Eq.~(\ref{eq:linestrength1}) is given in \cite{Tran2013}. $W(\omega_\text{BBR},T_{\text{BBR}})$ denotes the BBR energy distribution function at frequency $\omega_\text{BBR}$, which is given by:
\begin{equation}
W(\omega_\text{BBR} ,T_{\text{BBR}})=\frac{\hbar \omega_\text{BBR}^3}{\pi^2 c^3} \left( \text{e}^{\frac{ \hbar \omega_\text{BBR}}{k_B T_{\text{BBR}}}}-1 \right)^{-1}.
\end{equation}
Since the typical frequency of the internal degrees of freedom ($>$ 1~THz) differs from that of the external degrees of freedom ($<$ 1~MHz) by many orders of magnitude, any energy transfer mechanism between them must be of extremely high order and consequently negligibly small. Laser cooling of the external degrees of freedom may therefore be expected to have no significant effect on the temperature of the internal degrees of freedom, which are coupled strongly to (and in equilibrium with) the BBR field~\cite{Koelemeij2007b}.

We use \textsc{Mathematica} to solve Eq.~(\ref{eq:rateeq}) in order to obtain $\boldsymbol{\rho}(t)$. We subsequently find the relative \HD\ loss, $\epsilon$, by summing over the hyperfine state populations (62 hyperfine states in $(v,L)=(0,2)$) before (\textit{i.e.} $t=0$) and after ($t=10$) REMPD and computing:\\
\begin{equation}\label{eq:epsilon}
\epsilon = \frac{N_i-N_f}{N_i}= \frac{ \sum_{j=1}^{62} \rho_{j}(0)-\rho_{j}(10)}{ \sum_{j=1}^{62} \rho_{j}(0)}.
\end{equation}
Here $N_i$ and $N_f$ are the numbers of \HD\ ions present in the trap directly before and after REMPD, respectively. For a thermal ensemble of HD$^+$ ions, $\epsilon$ is a function of the variables $\nu-\nu_{0,\text{fit}}, T_{\text{HD}^+}$, and $I_{\text{laser}}$. In what follows we furthermore assume that $T_{\text{BBR}}=300$~K, which is the average temperature in our experimental setup.

\begin{figure}
\centering
\includegraphics[trim=0cm 0cm 13cm 12cm, clip, width=8cm]{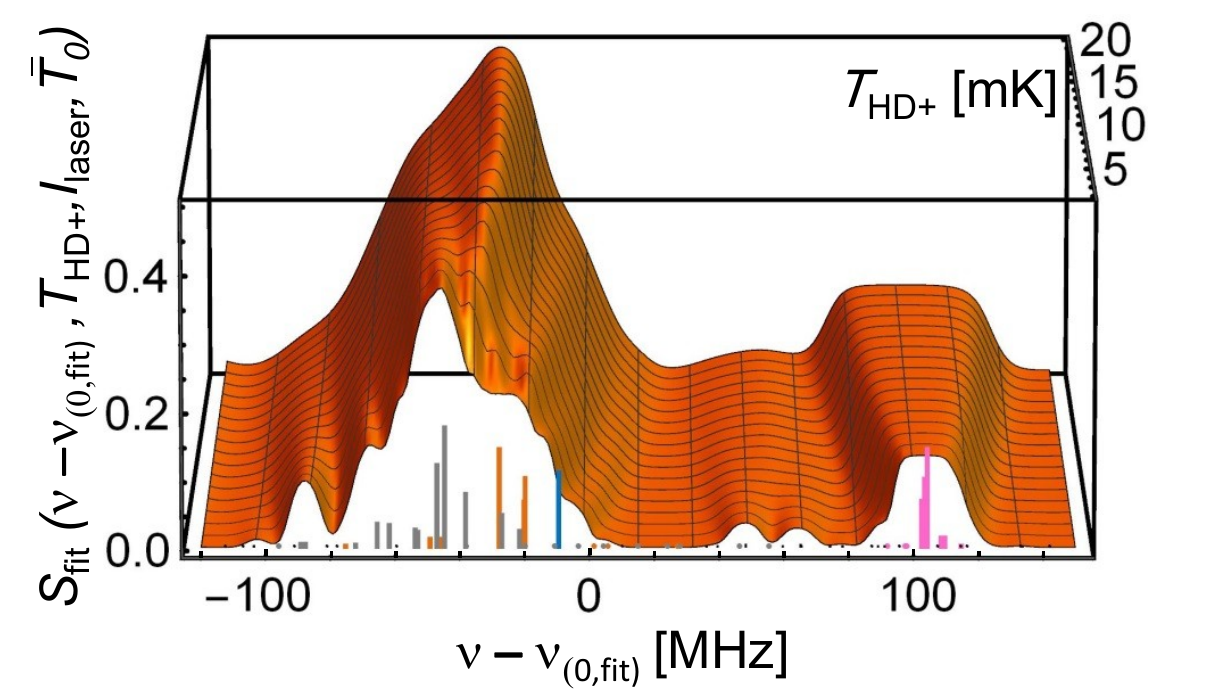}
\caption{\label{Sfitfigure}A plot of a the fit function $S_{\textrm{fit}}$ with $I_{\text{laser}}$=1.0$\times$10$^7$ Wm$^{-2}$, \Tzz=4~K and $\nu_{0,\text{fit}}=\nu_{\text{th}}$. On the frequency axis, the theoretical values of transitions between the particular hyperfine levels are depicted as sticks. The heights of the sticks correspond to their linestrengths. The stick colors are the same as those used in Fig.~\ref{REMPDabc} and distinguish different groups of transitions with similar quantum numbers $F$ and $S$. For decreasing HD$^+$ temperature, Doppler broadening is reduced and the hyperfine structure becomes more resolved. Effects of saturation are also visible in the spectrum.}
\end{figure}

The question arises what the relation is between the signal $S$ defined in Eq.~(\ref{S}) and the fractional loss $\epsilon$ defined above. In previous work it was assumed that $S$ and $\epsilon$ are interchangeable~\cite{Roth2006,Koelemeij2007a,Schneider2010,Koelemeij2012}. In Appendix~\ref{nlfunction} we study the dependence of the signal $S$ on the initial number of ions $N_i$ and the dissociated fraction $\epsilon$ in detail using realistic molecular dynamics (MD) simulations. We find that the fraction $\epsilon$ (which is a theoretical construct) can be mapped to the signal $S$ by use of a slightly nonlinear function,
\begin{equation}\label{Sfit}
S_{\text{fit}} = f_{\textrm{NL}}(\Tz,\epsilon)
\end{equation}
where \Tzz\ is defined as the effective \Be\ temperature along the $z$-axis during the initial secular scan of the REMPD cycle (see Appendix~\ref{nlfunction}). %The nonlinear relation $f_\text{NL}$ is derived and explained in detail in appendix \ref{nlfunction}.\\
This means we have to use a five-dimensional fit function
%%\begin{widetext}
\begin{equation}\label{Sfit2}
\begin{split}
&S_{\text{fit}}(\Tz,\epsilon(\nu-\nu_{0,\text{fit}}, T_{\text{HD}^+}, I_{\text{laser}})) \\ &=S_{\text{fit}}(\nu-\nu_{0,\text{fit}}, T_{\text{HD}^+},I_{\text{laser}}, \Tz).
\end{split}
\end{equation}
%%\end{widetext}
An analytical solution of the fit function proves difficult to find, whereas a numerical implementation of the fit function takes excessively long to compute. Therefore we compute values of $S_{\text{fit}}$ on a sufficiently dense grid of values $\nu-\nu_{0,\text{fit}}$ (encompassing 270 values between $-120$ and 150~MHz), $T_{\text{HD}^+}$ (20 values between 1 and 20~mK), $I_{\text{laser}}$, and $\Tz$ (9 values between $0.65\times 10^7$ and $2.1\times 10^7$~W$\,$m$^{-2}$), which we interpolate (again using \textsc{Mathematica}) to find a fast, continuous and smooth approximation to the function $\epsilon$, $\epsilon'$, which we insert into the function $S_{\text{fit}}$. The result is a five-dimensional (5D) fit function which is suitable for nonlinear least squares fitting. A 3D projection of $S_{\text{fit}}$ assuming fixed values of $I_\text{laser}$, \Tzz\ and $\nu_{0,\text{fit}}$ is plotted in Fig. \ref{Sfitfigure}.
The reason for treating $I_{\text{laser}}$ as a fit parameter instead of inserting a single fixed value is the following. The entire spectroscopy measurement is divided into 15 sessions each taken at a different day. To ensure reproducible laser beam intensities from session to session, we used diaphragms to overlap all laser beams with the 313~nm cooling laser, which itself is aligned with the \Be\ Coulomb crystal visually using the EMCCD camera. Using a mock version of this setup, we verified that this procedure leads to beam pointing errors up to 40$~\mu$m. Assuming a Gaussian distribution of beam pointing errors, we find an intensity at the location of the \HD\ ions which varies from the intensity in the center of the beam by a factor of 0.6 to 1. Since the spectral line shape is strongly saturated, these variations of the intensity only lead to small signal changes. It is therefore allowed to treat $I_{\text{laser}}$ as a free fit parameter which represents the average 782~nm laser intensity for all data points. Similarly, the variables $T_{\text{HD}^+}$  and $\Tz$ cannot be determined accurately \emph{a priori} and are treated as free fit parameters as well.

\subsection{Estimation of absolute trapped ion numbers}\label{absN}
As explained in Sec.~\ref{chemistry}, effects of chemistry in the Coulomb crystal significantly influence the measured lineshape of the \vv\ transition. The impact of such effects depends on (and can be estimated from) the absolute numbers of beryllium ions and molecular hydrogen ions. In order to estimate absolute numbers we combine results from MD simulations and spectroscopy. Similar as observed by Blythe \emph{et al.}~\cite{Blythe2005}, our mixed-species ion ensembles contain not only \Be\ and \HD, but also BeH$^+$, BeD$^+$, \HHD, and \HDD. In this paragraph we focus on the latter two species which are created during the \HD\ loading procedure through the exothermic reactions
\begin{equation}\label{m4m5reactions1}
\textrm{HD} + \textrm{HD}^+ \longrightarrow \textrm{H}_2\textrm{D}^+ + \textrm{D}
\end{equation}
and
\begin{equation}\label{m4m5reactions2}
\textrm{HD} + \textrm{HD}^+ \longrightarrow \textrm{HD}_2^+ + \textrm{H}.
\end{equation}
After loading, the excess HD gas is pumped out of the vacuum chamber within a few seconds, after which the background vapor resumes its steady-state composition (predominantly H$_2$). Reactions with HD therefore only play a significant role during and just after loading.

\begin{figure}
\centering
\includegraphics[width=6.5cm]{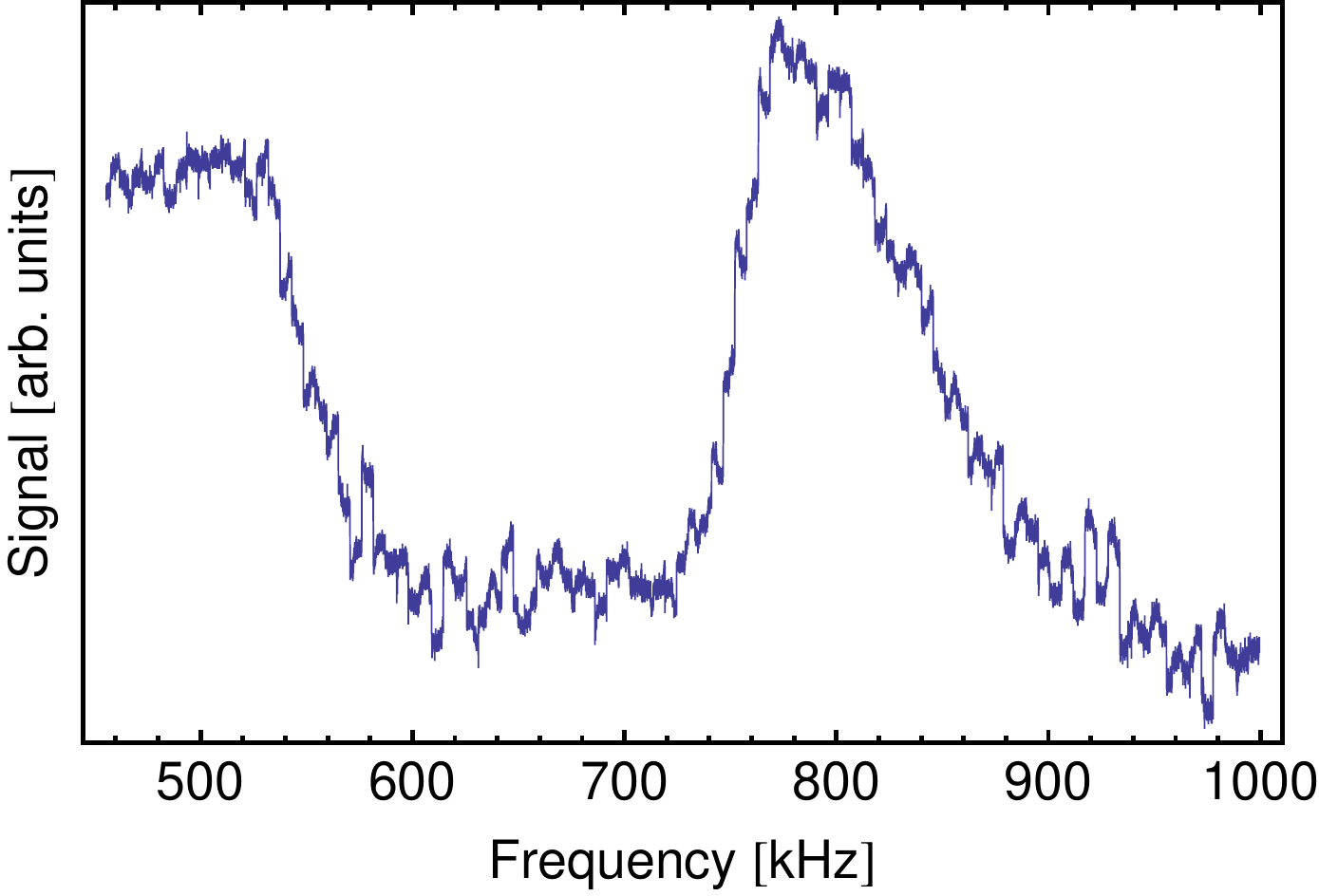}% Here is how to import EPS art
\caption{\label{SSm4m5ref}Fluorescence at 313~nm recorded by the PMT during a secular scan from 450 to 1000 kHz. The peak at 800 kHz indicates the presence of trapped particles with charge-to-mass ratio 1:3 (mostly HD$^+$). The fluorescence signal at 500 kHz indicates the presence of particles with charge-to-mass ratios of 1:4 and 1:5 which are attributed to \HHD\ and \HDD. Note that peaks belonging to the two species overlap and are not resolved.}
\end{figure}

\begin{figure}
\centering
\includegraphics[width=\columnwidth]{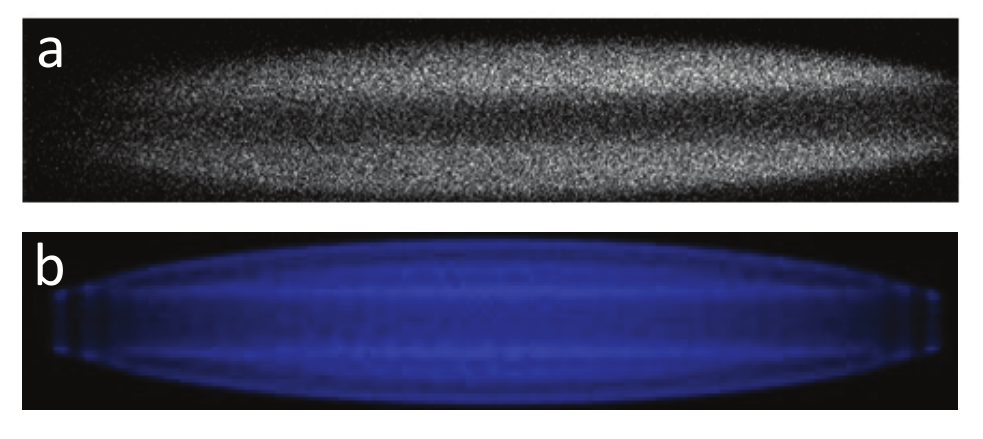}
\caption{\label{SimEMCCDions}(a) An EMCCD image of the \Be\ fluorescence just before a secular scan. (b) An image of ions obtained from a MD simulation based on 750 \Be\ ions and 50 singly-charged ions with masses 3, 4 and 5 amu.}
\end{figure}

Triatomic hydrogen ions can be detected by secular excitation. An example is shown in Fig.~\ref{SSm4m5ref}, where the peak in the PMT signal at the left is attributed to the overlapping peaks belonging to species with charge-to-mass ratios 1:4 and 1:5, corresponding to \HHD\ and \HDD, respectively.
In rf traps, lighter species experience stronger confinement. This is evident from EMCCD camera images which exhibit fluorescing \Be\ ions surrounding a dark core of lighter species (Fig.~\ref{SimEMCCDions}). The size of the core reflects the total number of light ions. We analyze this by means of MD simulations (Appendix~\ref{MDsim}), from which a relation is obtained between the size of the dark core and the number of trapped ions.
By comparing simulated and real EMCCD images, an average number of $\sim 750$ \Be\ ions is obtained. From the analysis of the image intensity profiles, we cannot distinguish the \HD\ from the triatomic molecular species. To solve this we use a collection of 140 EMCCD images taken before and after REMPD while the 782~nm laser was tuned at the same fixed frequency near the maximum of the \vv\ spectrum. From the lineshape model we estimate that the average relative \HD\ loss is 0.57 at this frequency. By comparing the EMCCD images taken before and after REMPD with the simulated images, we also determine the total initial and final numbers of light ions. Combining this with the expected loss of 57~\% of the \HD\ ions, we infer the ratio of \HD\ numbers to heavier molecular species (\HHD, \HDD). An average number of 43 \HD\ ions is obtained together with 60 ions of heavier species. We cannot determine the relative abundance of \HHD\ and \HDD, but previous observations indicate a branching ratio between Eqs.~(\ref{m4m5reactions1}) and (\ref{m4m5reactions2}) of 1:1 and, thus, equal abundances~\cite{Oka1992,Pollard1991}. The set of EMCCD images shows an appreciable spread in the size of the dark core and, in particular, the ratio of \HD\ to heavier triatomic hydrogen ions. Variations in both are due to uncontrolled shot-to-shot fluctuations of the HD background pressure during \HD\ loading. We find somewhat asymmetrical distributions of \HD\ ions and ions of heavier species, with a wider tail towards higher numbers. The resulting standard deviation of the number of \HD\ is 41 ions. This also indicates that analysis of EMCDD images (under the present conditions) is not suitable to replace the signal obtained by mass-selective secular excitation (Eq.~(\ref{S})).

For the treatment of effects of chemistry on the lineshape in Sec.~\ref{chemistry} below, we consider two scenarios:
 \begin{itemize}
\item Scenario a: $N_{\text{HD}^+}=43$, $N_{\text{H}_2\text{D}^+}=N_{\text{HD}_2^+}=30$
\item Scenario b: $N_{\text{HD}^+}=84$, $N_{\text{H}_2\text{D}^+}=N_{\text{HD}_2^+}=60$,
\end{itemize}
where scenario b reflects the one sigma upper variation (which well represents the width of the upper tail of the ion number distribution).

\subsection{Effect of micromotion}\label{mm}
The rf quadrupole field of the ion trap inevitably leads to micromotion of ions with non-zero displacement from the trap $z$-axis. In addition, excess micromotion may be caused by unwanted rf fields arising from geometrical imperfections of the trap electrode structure or phase differences between the rf electrodes~\cite{Berkeland1998}. In an ideal linear rf trap, micromotion is strictly radially oriented, but small imperfections in the trap geometry can cause excess micromotion with a component  along the trap axis and laser direction, thus adding phase-modulation sidebands to each hyperfine component in the \vv\ spectrum. Due to the combination of an asymmetric and saturated lineshape of the \vv\ spectrum, these sidebands can lead to a shift of $\nu_0$. Therefore the micromotion amplitude along the 782~nm laser needs to be determined. As the laser propagates virtually parallel to the trap axis, and since the \HD\ ions are always located near the trap axis, we are primarily concerned with the possible axial micromotion component.

The HD$^+$ axial micromotion amplitude $x_{\text{HD}^+}$, can be determined through fluorescence measurements of a trapped string of beryllium ions using a modified version of the photon-rf field correlation technique~\cite{Berkeland1998}.  The idea here is to radially displace a string of about 10 \Be\ ions by $\sim$100~$\mu$m by applying a static offset field. This will induce significant radial micromotion, in addition to the axial micromotion. The 313~nm cooling laser propagates at an unknown but small angle $\theta$ ($\theta<$10~mrad) with respect to the trap axis, and may therefore have a nonnegligible projection along the radial direction. In Appendix~\ref{MMfitfunction} we show that if the rf voltage amplitude, $V_0$, is varied, the axial micromotion amplitude scales linearly with $V_0$, while the radial one varies as $\theta/V_0^2$. The latter behavior stems from the $V_0$-dependent confinement and the concomitant variation of the \Be\ radial displacement with $V_0$. Thus, measuring the micromotion amplitude for various values of $V_0$ allows separating the radial and axial contributions.

\begin{figure}
\centering
\includegraphics[width=6.5cm]{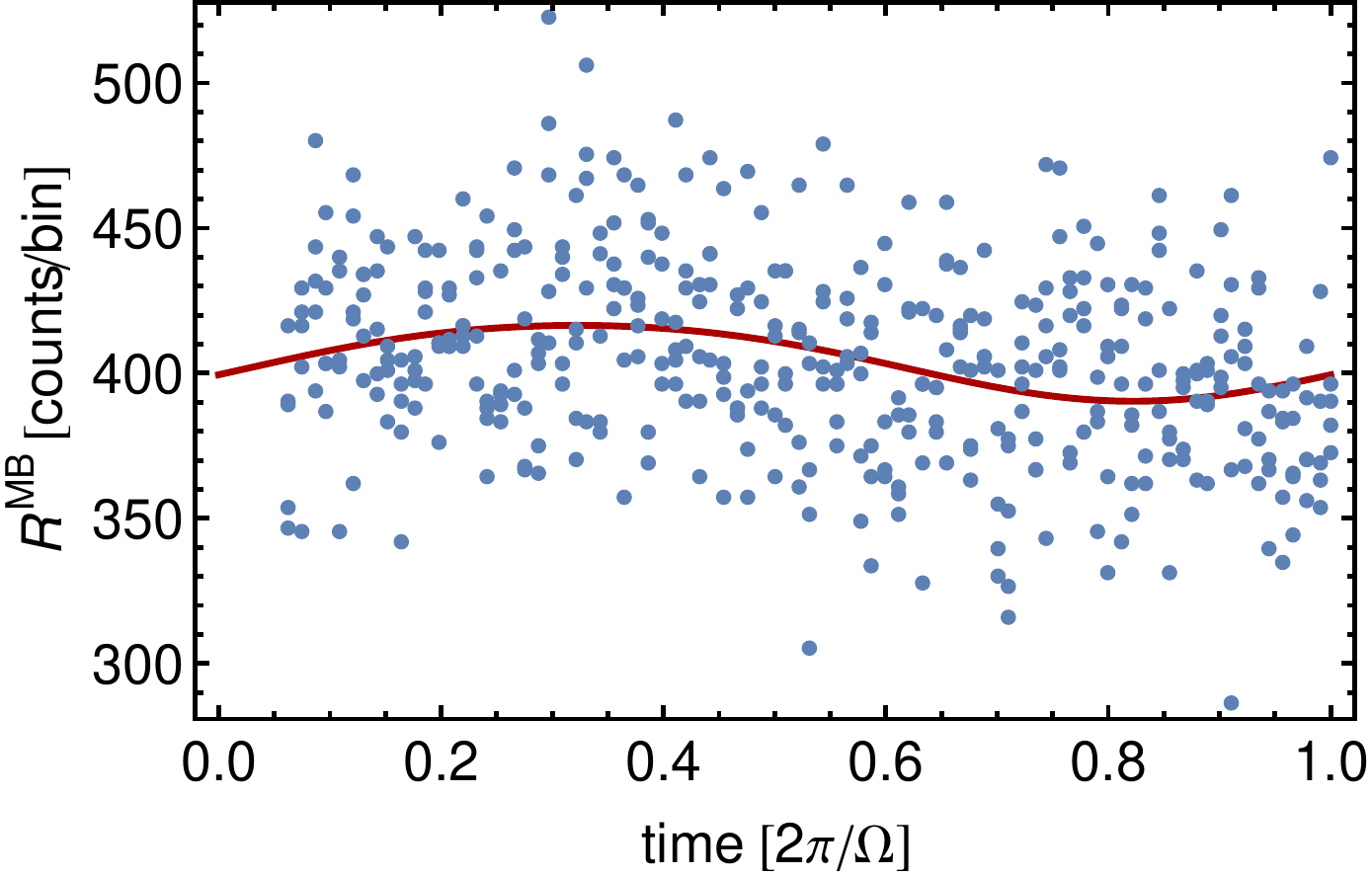}
\caption{\label{mmbins}Micromotion-induced \Be\ fluorescence modulation detected with the TAC setup. The data are fitted using Eqs.~(\ref{eq:MMR}) and (\ref{eq:MMkdotv}).}
\end{figure}

To determine the micromotion amplitude, we use a similar setup as described in~\cite{Berkeland1998}. Photons detected with the PMT are converted to electrical pulses and amplified by an amplifier-discriminator, which generates a START pulse at time $t_i$. Subsequently, a STOP pulse is generated at time $t_f$ at the first downward zero crossing of the rf signal. A time-to-amplitude converter (TAC) converts the duration between the START and STOP pulses to a voltage. We record the TAC output voltage with a digital phosphor oscilloscope for 400~ms. We subsequently process the stored voltage trace with a computer algorithm to obtain a histogram of START-STOP time delays, employing 1-ns bins in a range 0-76~ns (\textit{i.e.} one rf cycle). The bin heights thus reflect the scattering rate as a function of the rf phase, and micromotion will lead to a modulation of the scattering rate about its mean value (Fig.~\ref{mmbins}). The Be$^+$ scattering rate (indicated as $R^{\text{MB}}$, where MB stands for Maxwell-Boltzmann) is given by:
%%\begin{widetext}
\begin{equation}\label{eq:MMR}
\begin{split}
&R^{\text{MB}}= \frac{\Gamma}{2}\sqrt{\frac{m_{\text{Be}}}{2\pi k_B T}} \times \\ &\int\frac{I/I_\text{sat}}{I/I_\text{sat}+1+(2(\Delta-\mathbf{k}\cdot (\mathbf{v_{\mu}} - \mathbf{v}))/ \Gamma)^2}\, \text{e}^{ -\frac{m_{\text{Be}}v^2}{2k_B T} } dv
\end{split}
\end{equation}
%%\end{widetext}
where $m_{\text{Be}}$ is the \Be\ mass, $T$ the \Be\ temperature, $I$ the 313~nm laser intensity, $I_{\text{sat}}$ the saturation intensity of the 313~nm cooling transition in Be$^+$, $\Delta = 2 \pi \times - 25$~MHz the detuning of the 313~nm laser light, $\mathbf{k}$ the wavevector of the 313~nm laser, and $\mathbf{v}$ and $\mathbf{v}_\mu$ the secular and micromotion velocities of the \Be\ ions. While the rf voltage is being varied, the 313~nm laser is displaced so that the ions always are at the maximum of the Gaussian laser intensity profile. The $\mathbf{k}\cdot \mathbf{v_{\mu}}$ term can be written as
\begin{equation}\label{eq:MMkdotv}
\mathbf{k}\cdot \mathbf{v_{\mu}}=k x_{0,k} \Omega \sin(\Omega (t-t_0)),
\end{equation}
where $x_{0,k}$ is the amplitude of \Be\ micromotion along the direction of the laser wavevector and $t_0$ is a time offset.
We extract $x_{0,k}$ by fitting Eq.~(\ref{eq:MMR}) to the acquired fluorescence histogram. Repeating this procedure for various values of $V_0$, a list of data points of the form $(V_0, x_{0,k})$  is obtained. We subsequently extract the radial and axial micromotion components by fitting a model function to these data. The model function is derived in Appendix~\ref{MMfitfunction}.

\begin{figure}
\centering
\includegraphics[width=6.6cm]{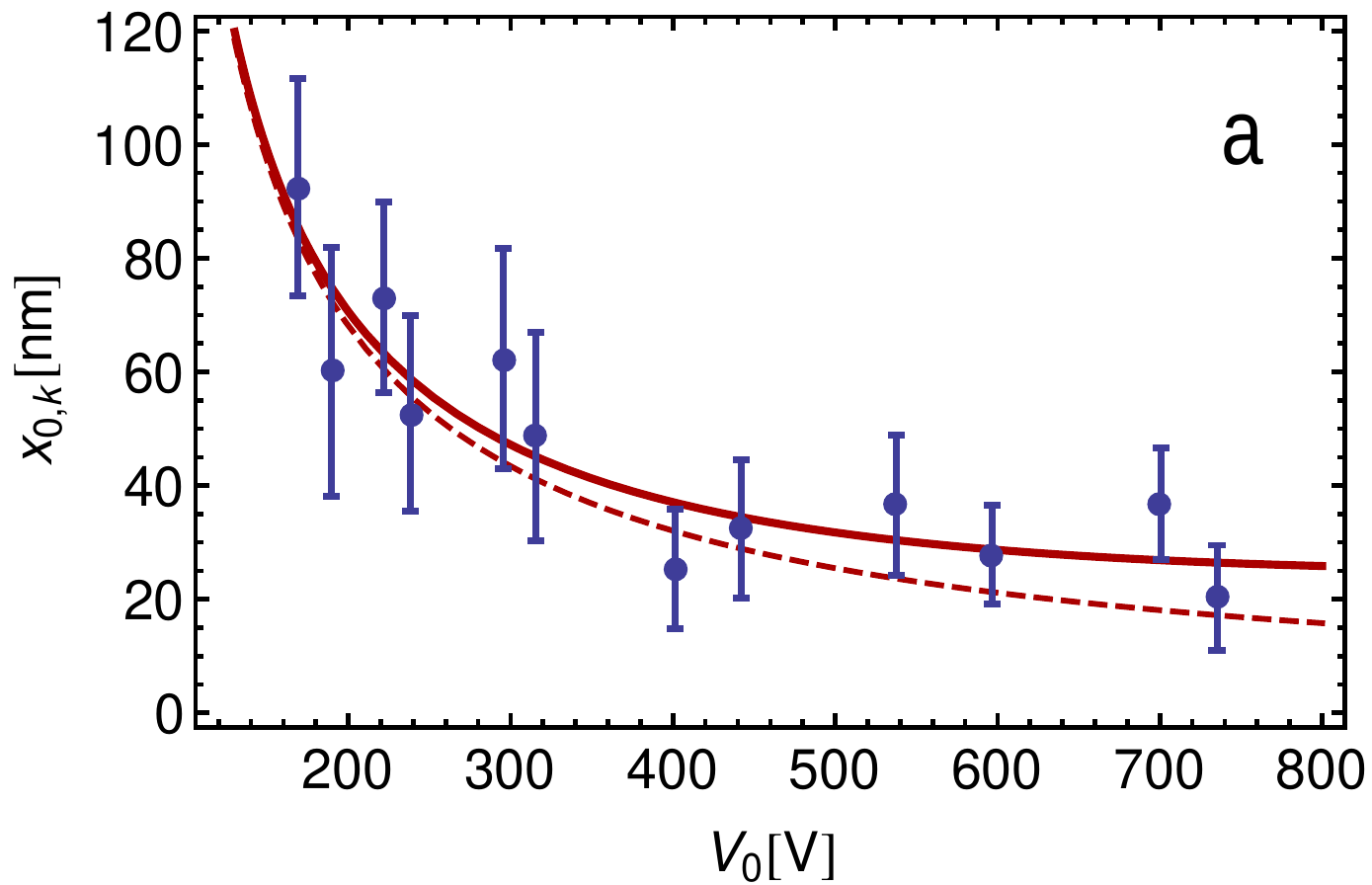}
\includegraphics[width=6.5cm]{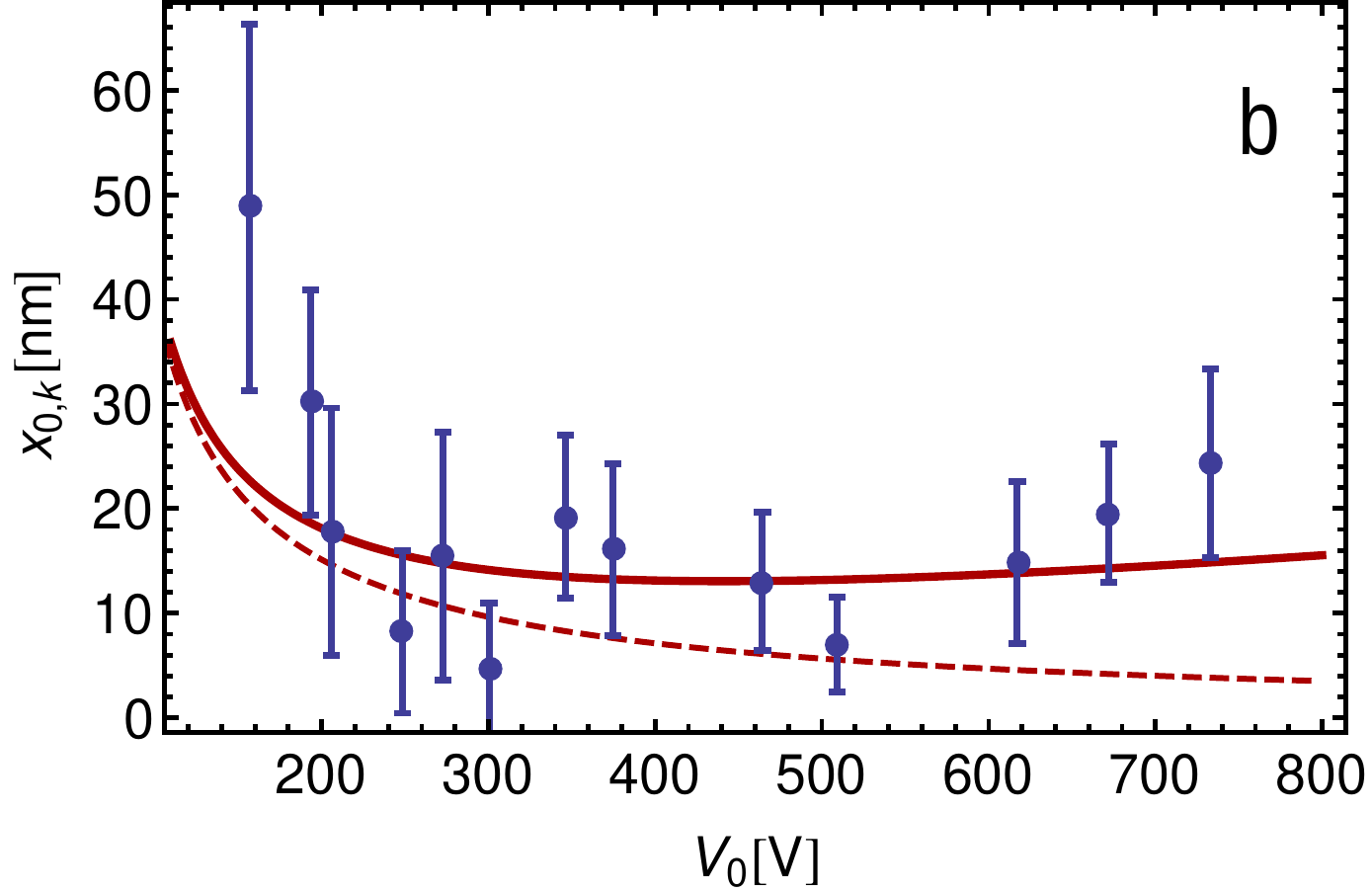}
\caption{\label{mmvdy}\Be\ micromotion amplitude along the 313~nm wavevector versus rf voltage amplitude for an offset in the horizontal (a) and vertical (b) direction. The red solid curve is the model fit function [Eq.~(\ref{xHDthetaexp})] which includes $x_{\text{HD}^+}$ as a fit parameter. For comparison the fitted curve with $x_{\text{HD}^+} =0$ is also shown (dashed curve).}
\end{figure}

The procedure of displacing a string of Be$^+$ ions and varying $V_0$ is carried out for both the horizontal and vertical directions. The data and fit functions are shown in Fig.~\ref{mmvdy}
and an average axial micromotion amplitude $x_{\text{HD}^+}$ (the projection along the 782~nm wavevector) of 11(4)~nm is found. As explained in Appendix~\ref{MMfitfunction}, the radial micromotion contribution (due to a possible small angle of the 782~nm laser with the trap axis) averages to zero.

We incorporate the micromotion effect by extending the lineshape function $D_z$ in Eq.~(\ref{eqMrempd}) with sidebands at frequencies $m \Omega$ with amplitudes $J_m^2 (k_{782} x_{\text{HD}^+})$. Here, $k_{782}$ denotes the wavevector of the 782~nm laser, and the $J_m$ are Bessel functions of the first kind, with $m$ an integer in the range $[-3,3]$.

\subsection{Effects of chemistry in the Coulomb crystal}\label{chemistry}
During REMPD, H$_2$ molecules in the background gas can react with the ions in the Coulomb crystal. Such reactions can be divided into two classes: (i) elastic collisions,  and (ii) inelastic collisions, during which a chemical reaction or charge exchange occurs, and chemical energy is converted into kinetic energy. In general, the kinetic-energy transfer to the ion from elastic collisions with room-temperature particles is much lower than the chemical energy released from inelastic collisions.

At close range $r$, the electric field of the ion polarizes the neutral molecule which results in an attractive interaction potential $U(r)=-\alpha Q^2/(8 \pi \epsilon_0 r)^4$.  Here $\alpha$ denotes the polarizability volume (in m$^3$)\footnote{The polarizability volume and the polarizability in SI units are related through $\alpha = \alpha_\text{SI}/(4 \pi \epsilon_0$)} of the molecule, $\epsilon_0$ is the electric constant, and $Q$ is the elementary charge. If we integrate the interaction force over the trajectory of the neutral near the ion (assuming a relatively large impact parameter $b > b_{\text{crit}}$; see below), we obtain a change of momentum which corresponds to a velocity kick of tens of meters per second for most species.

If a neutral atom or molecule and an ion approach each other within a critical impact parameter $b_{\text{crit}}=(\alpha q^2/\pi \epsilon_0 \mu_{\text{red}} v)^{1/4}$, where $\mu_{\text{red}}$  and $v$ are the reduced mass and relative velocity of the pair, a so-called Langevin collision occurs, during which the collisional partners spiral towards each other and a chemical reaction can occur at very short range~\cite{Hasted1964}. The chemical reaction products contain hundreds of meV of kinetic energy, which is dissipated into the ion crystal which itself only contains about 2~meV of kinetic energy (at 10~mK). A possible adverse side effect is that such collisions may lead to time-averaged velocity distributions which deviate from thermal distributions. Table~\ref{tab:chemreactions} shows the relevant reactions during REMPD along with the released chemical energy and their reaction rates.
\begin{table*}
\small
\caption{\label{tab:chemreactions}%
Chemical reactions occurring in the Coulomb crystal during REMPD. Rates are computed for the average H$_2$ background pressure observed during the measurements.}
%%\begin{ruledtabular}
%%\begin{tabular}{lllll}
\begin{tabular*}{1\textwidth}[t]{@{\extracolsep{\fill}}llll}
\hline
\hline
\textrm{Number}&\textrm{Reaction}&\textrm{Energy of ionic product (eV)}&\textrm{Rate(s$^{-1}$)}\\
%%%\colrule
\hline
1&\HD\ + $h\nu+h\nu' \longrightarrow$ D$^+$ + H & 0.41 & 0-10$^\text{a}$ \\%\footnote[1]{The rate (in $s^{-1}$ per molecule) of D$^+$ production is dependent on time and frequency in the \vv\ spectrum.}\\
2&\HD\ + H$_2$ $\longrightarrow$ \HHD\ + H & 0.36 & 0.0042\\
3&\HD\ + H$_2$ $\longrightarrow$ \HHH\ + D &0.66  & 0.0014 \\
4&\HHD\ + H$_2$ $\longrightarrow$ HD + \HHH  & 0.016 & 0.0019 \\
5&\HDD\ + H$_2$ $\longrightarrow$ D$_2$ + \HHH & 0.017& 0.0004  \\
6&\HDD\ + H$_2$ $\longrightarrow$ HD + \HHD\ & 0.022 & 0.0015  \\
7&\Be($^2$P$_{3/2}$) + H$_2$ $\longrightarrow$ BeH$^+$ + H & 0.25 &0.0019/0.005$^\text{b}$ \\% \footnote{The rate is dependent on the fraction of time a \Be\ ion spends in the excited $^2$P$_{3/2}$ state, which is dependent on the 313 nm laser intensity and detuning ($-80$~MHz or $-300$~MHz, respectively).} \\
\hline
\hline
\end{tabular*}
\begin{tabular*}{1\textwidth}[t]{@{\extracolsep{\fill}}p{1\textwidth}}
{\begin{footnotesize}$^\text{a}$The rate (in s$^{-1}$ per molecule) of D$^+$ production is dependent on REMPD time and frequency of the 782~nm laser. \end{footnotesize}}  \\
{\begin{footnotesize}$^\text{b}$The rate is dependent on the fraction of time a \Be\ ion spends in the excited $^2$P$_{3/2}$ state, which is dependent on the 313 nm laser intensity and detuning ($-80$~MHz or $-300$~MHz, respectively).\end{footnotesize}}
\end{tabular*}
%%\end{ruledtabular}
\end{table*}
The reaction numbered 1 corresponds to the REMPD process itself. From observations reported in~\cite{Esry1999} we infer that the ratio of \HD\ $\longrightarrow$ D$^+$ + H and \HD\ $\longrightarrow$ H$^+$ + D is approximately 1:1. The charge-to-mass ratio of H$^+$ is too large for this product to be stably trapped, but the D$^+$ ions can stay trapped and can orbit the Coulomb crystal for many seconds. The reaction rate of 1 is calculated from the REMPD model described in Sec.~\ref{lsmodel}, and is dependent on the frequency of the 782~nm laser.

Reaction 7 occurs most frequently due to the large number of \Be\ ions present in the trap.  The reaction rate is obtained from the exponential decay of the measured 313~nm fluorescence emitted by a Coulomb crystal of \Be\ ions, and is in good agreement with the rate estimated from the Langevin cross section given the background pressure of $1\times 10^{-8}$~Pa in our apparatus~\cite{Wineland1997}. The different rate constants of reactions 2 and 3 illustrate the fact that \HD\ can react with H$_2$ in two ways: either the H$_2$ breaks apart, donating an H atom to the \HD\ molecule, or the \HD\ breaks apart, after which an H$^+$ or D$^+$ is added to the neutral molecule. According to~\cite{Oka1992} and~\cite{Pollard1991} the probability of each scenario is approximately 50~\%. In case the ion breaks apart, the probability that either a H$^+$ or a D$^+$ is donated to the H$_2$ molecule is also 50 \%. This leads to a ratio of reaction 2 to 3 of 3:1. The rate of reaction 2 can be measured (keeping in mind that HD$^+$ and H$_3^+$ in reaction 3 have the same charge-to-mass ratio and therefore cannot be distinguished) by applying the measurement scheme depicted in Fig.~\ref{fig:ssREMPDss} without 782~nm laser, which is further described in Sec.~\ref{bg}. The rates of reactions 4, 5 and 6 are obtained from~\cite{Giles1992}. The kinetic energies of the chemical products are calculated by using the binding energies and energy and momentum conservation laws.

\begin{figure}
\centering
\includegraphics[trim=0cm 0cm 13cm 13cm, clip, width=8cm]{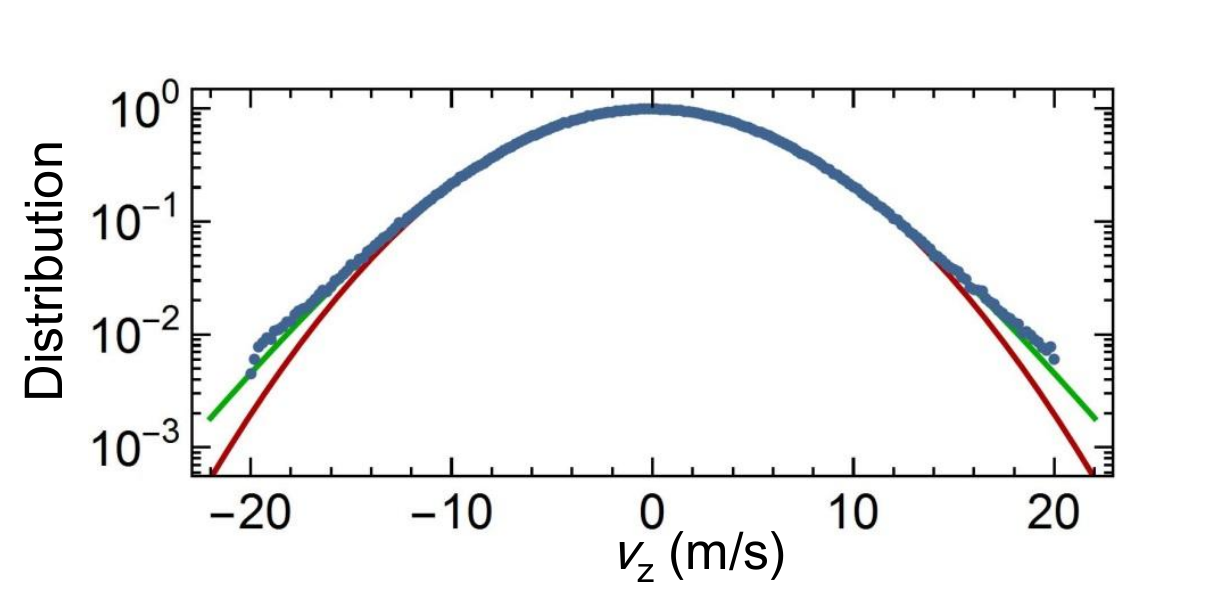}
\caption{\label{vdistrq}The $z$-velocity distribution of 43 \HD\ molecules obtained from 100~ms simulation. 14 D$^+$ ions with an initial velocity of 6300 ms$^{-1}$ heat up the ion crystal and give rise to a velocity distribution which differs slightly from a Gaussian (red curve, fitted temperature 11.60(3) mK)) and is better described by a $q$-Gaussian (green curve, fitted temperature 11.00(3)~mK, and fitted $q$ parameter 1.070(3)).}
\end{figure}

MD simulations show that the fast ionic chemical products may heat up the Coulomb crystal by 1--2~mK (depending on the REMPD rate), and that the \HD\ velocity distribution becomes slightly non-thermal. A MD simulation of a Coulomb crystal containing 750 laser-cooled \Be\ ions, 40 sympathetically cooled \HD\ ions and 14 fast D$^+$ ions produces the \HD\  $z$-velocity distribution shown in Fig.~\ref{vdistrq}.
Details of the MD code are given in Appendix~\ref{MDsim}. It turns out that the $z$-velocity distribution deviates from a Gaussian curve and is better described by a $q$-Gaussian~\cite{Umarov2008}, which is a Gaussian curve with higher wings parameterized by a parameter $q$. For $1<q<3$, the $q$-Gaussian is written as
%%\begin{widetext}
\begin{equation}
Q(\omega,\beta,q)=\frac{\sqrt{q-1}\left( 1+ \frac{(q-1)(\omega_0-\omega)^2}{2\beta^2}\right)^\frac{1}{1-q}\Gamma\left(\frac{1}{q-1}\right) }{\sqrt{2 \pi}\ \beta \ \Gamma\left(-\frac{q-3}{2(q-1)}\right)}.
\end{equation}
%%\end{widetext}
Here, $\omega$ and $\omega_0$ are the frequency and center frequency, $\Gamma$ is the gamma function, and $\beta$ is analogous to the standard deviation of a Gaussian distribution, which is related to the Doppler width. For $q=1$, the function reduces to a regular Gaussian distribution. While we use $q$-Gaussians to describe the simulated data, the observation that $q$-Gaussians describe the simulated velocity distributions well is merely empirical, and we did not derive this velocity distribution from a particular physical model. Since the value $\nu_{0,\text{fit}}$ turns out to be sensitive to the shape of the velocity distribution, it is important that we insert the lineshape based on the correct velocity distribution in Eq.~(\ref{eqMrempd}), and specify the bounds to within this distribution is valid. An initial analysis reveals that $\nu_{0,\text{fit}}$ may shift by several hundreds of kHz by implementing a $q$-Gaussian with $q$ varying between 1.0 and 1.1. Note that another recent study of MD simulations independently confirmed the non-thermal character of velocity distributions of laser-cooled ion crystals due to collisions with background gas molecules~\cite{Rouse2015}.

In the remainder of this section, we determine the $q$-values applicable to our REMPD measurement with the help of MD simulations. The value of $q$ is dependent on the number of trapped fast ions, $N_{\text{fast}}$. The more fast ions, the higher the $q$-value. We note that $N_{\text{fast}}$ is frequency dependent (more D$^+$ are produced near the peak of the REMPD spectrum) as well as time dependent (the production rate of D$^+$ is governed by the rate equations, Eq.~(\ref{eq:rateeq})).

When a fast ion collides with a cold ion, each particle may undergo a nonadiabatic transition to a different solution of the Mathieu equation which governs its motion in the trap~\cite{Chen2013}. This implies that a fast ion can either lose or gain energy during such a collision. Since the energy change per collision is relatively small, fast ions can retain their high speeds in the trap for many seconds. Values for $N_{\text{fast}}$ can be obtained by solving the rate equation:
\begin{eqnarray}\label{Nfast}
\frac{\partial N_{\text{fast}}}{\partial t}= \alpha_{\text{prod}} N_{\text{source}}-\alpha_{\text{relax}}N_{\text{fast}}-\alpha_{\text{esc}}N_{\text{fast}}
\end{eqnarray}
The term $\alpha_{\text{prod}}$ is the production rate from a number of $N_{\text{source}}$ particles, such as \Be\ or \HD.  In Sec.~\ref{absN} we determined $N_{\text{source,Be}^+}$=750 and  $N_{\text{source,HD}^+} = 43$ or $N_{\text{source,HD}^+}= 84$. $\alpha_{\text{relax}}$ is the rate at which fast ions are cooled and become embedded within the Coulomb crystal, while $\alpha_{\text{esc}}$ is the rate at which (fast) ions escape from the trap. The values of $\alpha_{\text{prod}}$ correspond to the reaction rates in Table~\ref{tab:chemreactions}. However, obtaining $\alpha_{\text{relax}}$ and $\alpha_{\text{esc}}$ requires a multitude of individual MD simulations, with simulation periods of several seconds each. Even with current academic supercomputers, the total time to perform such simulations is prohibitively long.
Therefore, we consider two extreme scenarios for relaxation and escape rates of trapped fast ions:
\begin{itemize}
\item Scenario 1: A minimum number of fast ions is present in the trap. $\alpha_{\text{relax}}$ and $\alpha_{\text{esc}}$ are set to their maximum value of one per second for all species, which is based on the observation that no ion loss and no relaxation are observed over simulated times up to 800 ms. This scenario will produce the smallest value of $q$.
\item Scenario 2: A maximum number of fast ions is present in the trap, which is realized by setting $\alpha_{\text{relax}}$ and $\alpha_{\text{esc}}$ to their minimum value of zero. All fast ions remain in the trap at high speed for the entire 10 seconds of REMPD. This scenario leads to the largest value of $q$.
\end{itemize}
%%In Fig. \ref{fig:Nfastions} the solutions of Eq. \ref{Nfast} are shown as a function of time for BeH$^+$ and D$^+$ %for the two scenarios.
%%\begin{figure*}
%%\includegraphics[trim=0cm 0cm 0cm 0cm, clip, width=15cm]{Nfastions2}% Here is how to import EPS art
%%\caption{\label{fig:Nfastions}(Color online) (a-b) The predicted number of trapped fast BeH$^+$ molecules during the different measurement steps based on (a) scenario 1 and (b) scenario 2.  For (a) and (b), different colors  represent a ratios of number of particles that escape the trap to the number that relaxes within the Coulomb crystal. Green, yellow, blue correspond to the ratios 40:60, 50:50 and 60:40. (c-d) The predicted number of trapped fast D$^+$ particles during REMPD for scenario 1a (c) and scenario 2b (d). Different colors represent here the height of the REMPD signal, which is a measure for the amount of produced D$^+$ . Blue, yellow, green, red, purple correspond to REMPD signals of 0.1, 0.2, 0.3, 0.4, 0.5, all for a ratio escape:relax of 50:50.}
%%\end{figure*}
We make the assumption that fast ions do not mutually interact when present in numbers of ten or less, so that the observations based on MD simulations with ten fast ions are also valid for smaller numbers of fast ions. Note that BeH$^+$ ions are already created during the first secular scan before the REMPD phase starts. Fast specimen of \HHD\ and \HHH\ occur less frequently in the trap, and in line with the extreme scenarios introduced above we assume zero \HHD\ and \HHH\ for scenario 1, and 3 \HHD\ and 1 \HHH\ for scenario 2. Together with scenarios a and b described in Sec.~\ref{absN} this gives us four scenarios in total, and therefore four different spectral fit functions and four different $\nu_{0,\text{fit}}$ results.

\begin{figure}
\centering
\includegraphics[width=7.7cm]{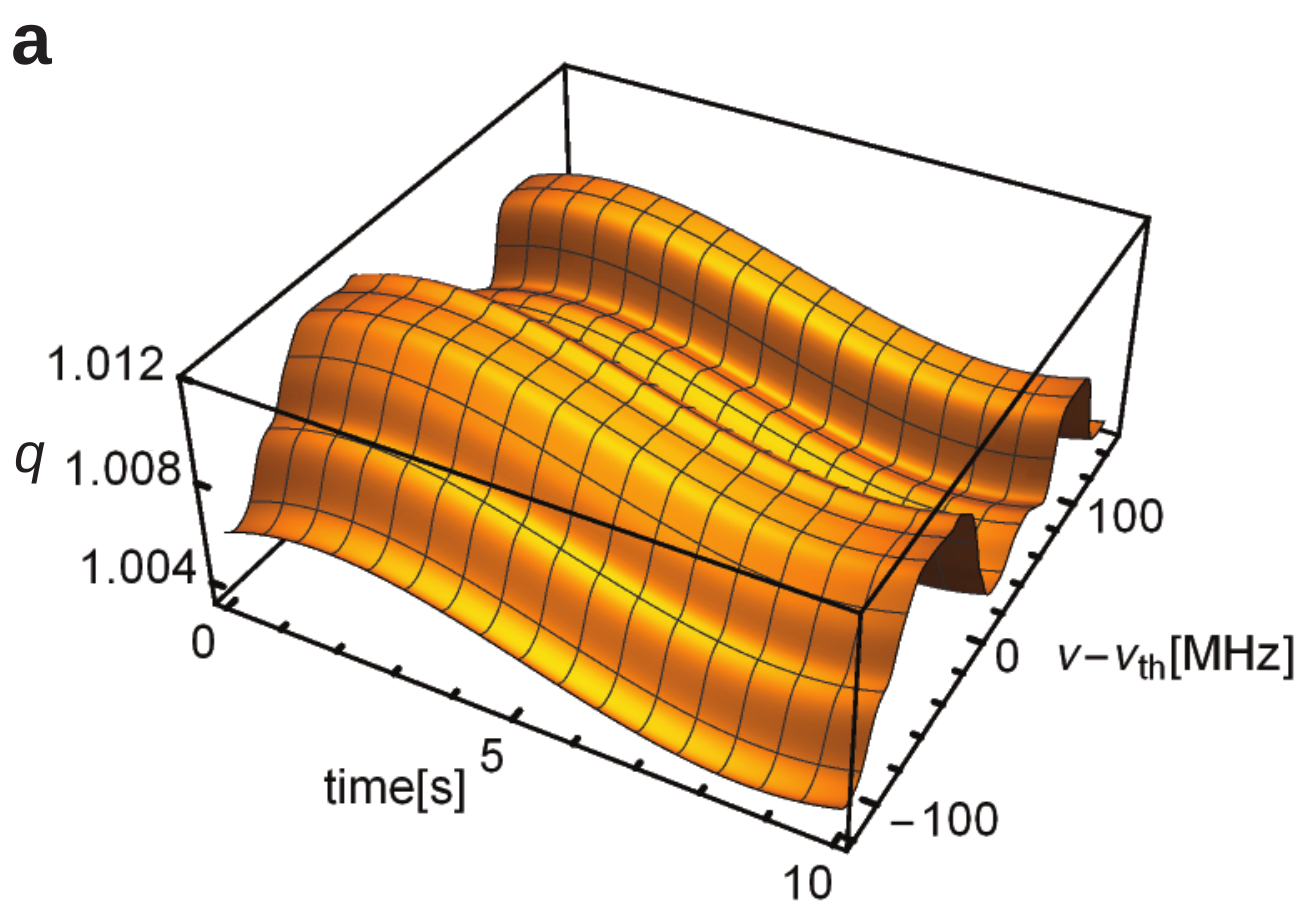}
\hfill
\centering
\includegraphics[width=7.8cm]{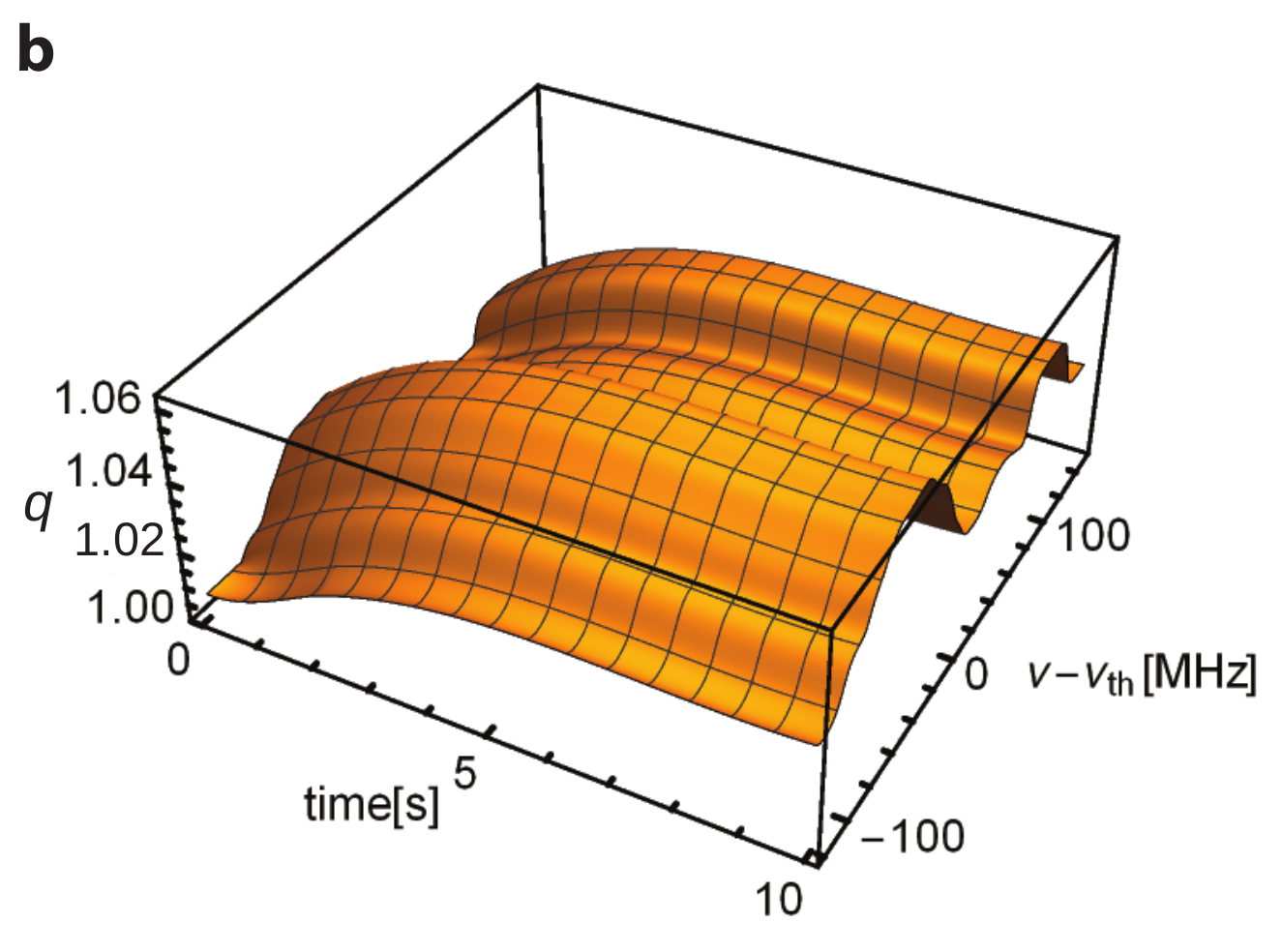}
\caption{\label{D3functionsq}3D $q(t,\nu)$ plots for scenario 1a (a) and 2b (b). Note that scenario 1a nearly corresponds to the case $q=1$, \textit{i.e} the case in which all chemistry-induced effects can be ignored. See text for further details.}
\end{figure}

For all possible combinations of fast ions present during REMPD (\textit{e.g.} 1 D$^+$, 2 BeH$^+$ and 1 \HHD\ or  3 D$^+$, 3 BeH$^+$ and 3 \HHD\ ) an MD simulation is carried out. From these simulations, the \HD\ $z$-velocity distribution is determined and a $q$-Gaussian is fitted which results in one $q$ value for each simulated case. As mentioned before, the production rate of fast D$^+$ depends on the REMPD rate (and thus on the 782~nm laser frequency $\nu$), which itself depends on the time-dependent number of available HD$^+$ ions in the target state. To take this properly into account we introduce a time and frequency dependent parameter  $q(t,\nu)$ as follows. For each of the four scenarios, the number of fast D$^+$ is simulated on a grid of different REMPD durations, $t_j$, and of different frequencies, $\nu_j$ of the 782~nm laser. On each point of this two-dimensional grid, the number of fast D$^+$ is combined with the number of other fast ions, and the corresponding value of $q(t_j, \nu_j)$ is looked up in a library of $q$ values, obtained from many MD simulations performed with various combinations and abundances of fast ion species. Interpolation of the grid $q(t_j, \nu_j)$ leads to a smooth continuous function $q(t,\nu)$, which is subsequently inserted into the lineshape function $D_z$ used in Eqs.~(\ref{eq:rateeq}) and (\ref{eqMrempd}). Figure~\ref{D3functionsq} shows the 3D plots of $q(t,\nu)$ for the different scenarios.

\begin{figure}
\centering
\includegraphics[width=7 cm]{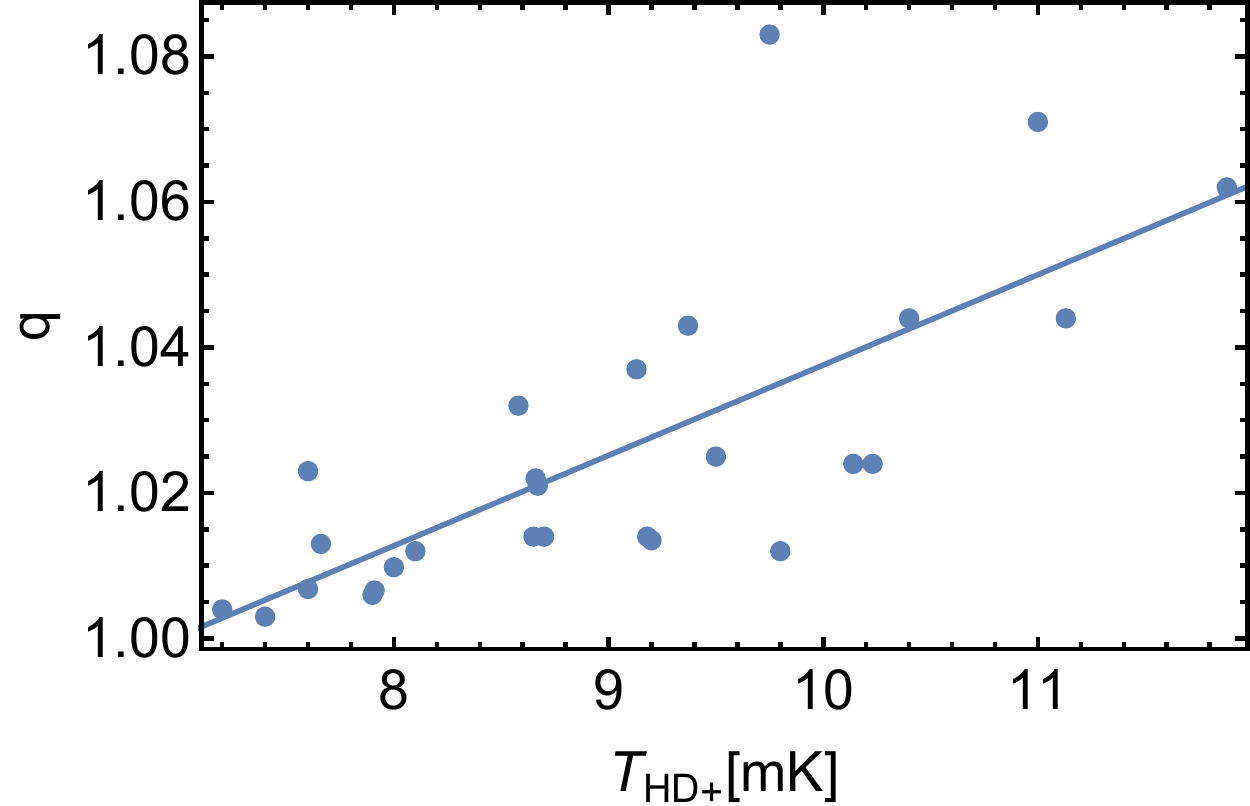}
\caption{\label{qTrelation}The relation between the temperature of a simulated Coulomb crystal and the $q$ value. Each point represents a simulation of approximately 100~ms. The data are best fitted with a linear function (blue line).}
\end{figure}

%Returning to the spectral analysis, those fit functions $q(t,\nu)$ are inserted in the as the q-parameters in the q-Gaussian lineshapes, which are subsequently inserted into formula \ref{eqMrempd}, giving us $\nu_0$.\\
Besides the $q$ value, also the ion temperature $T_{\text{HD}^+}$ is frequency and time dependent. Due to a larger number of D$^+$ ions at the top of the spectrum than at the wings, the temperature differences between top and wings can reach a few mK. We note that the increase of $q$ and $T_{\text{HD}^+}$ share the same origin (namely collisions with fast ions), and in Fig.~\ref{qTrelation} we show the relation between $q$ and $T_{\text{HD}^+}$, obtained from fitting
$q$-Gaussians to simulated velocity distributions. We find a linear dependence of the form $ T_{\text{HD}^+}\left[1+h (q(t,\nu)-1)\right]$, with the slope $h$ obtained from the fit. In scenario 1, the temperature difference between top and wings is found to be 0.5~mK. For scenario 2, the estimated temperature difference is 2.5(5)~mK, where the uncertainty of 0.5~mK is treated as one standard deviation. The $t$ and $\nu$ dependent temperature is also included in $D_z$.

\begin{figure}
\centering
\includegraphics[width=8.7cm]{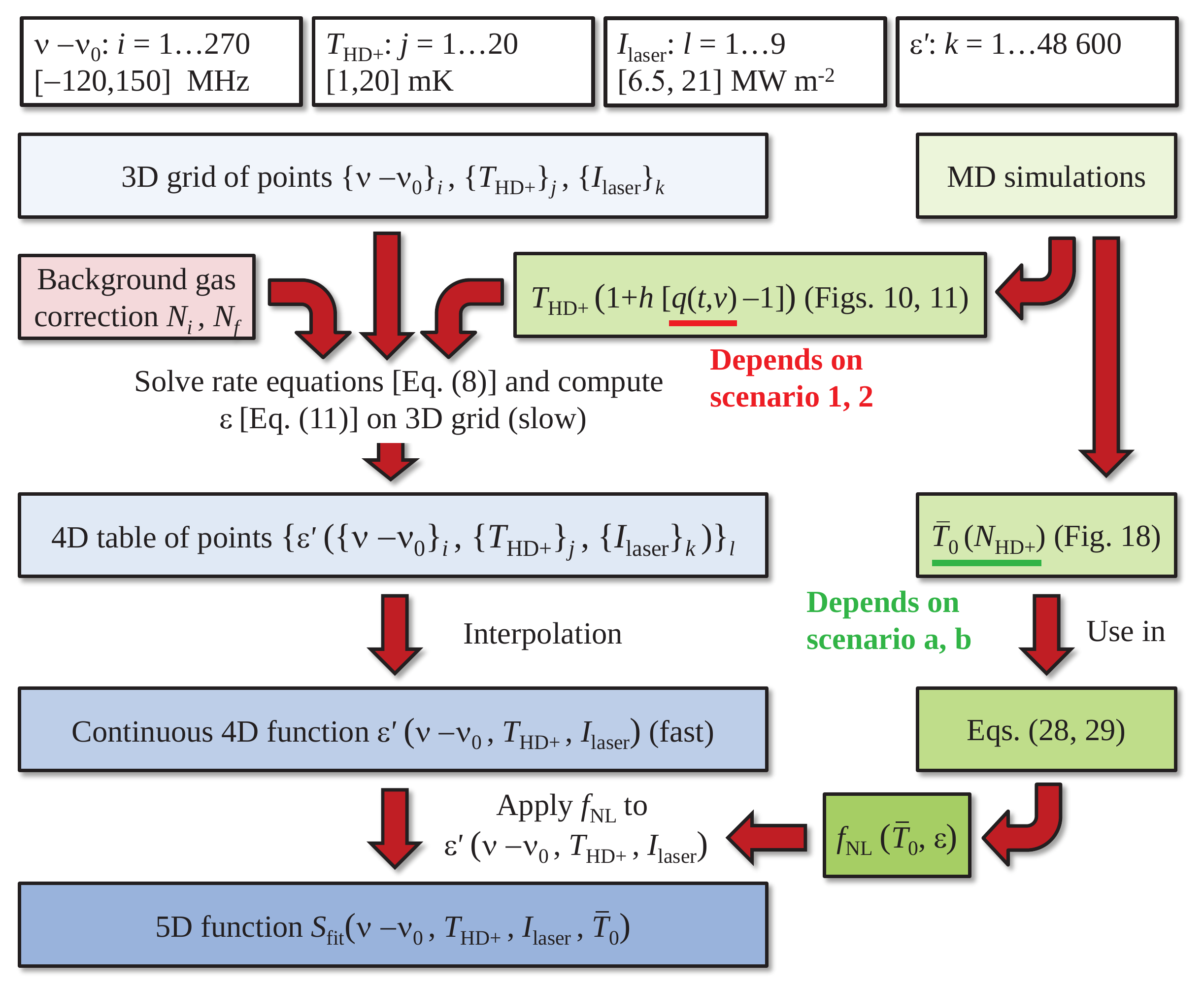}% Here is how to import EPS art
\caption{\label{fig:Sfitschematic} (Color online) Flow diagram showing the various steps involved in the construction of $S_\text{fit}$, and the role of the four scenarios considered here. As indicated at the top, the starting point is a 3D grid containing 48,600 entries, on which the rate equations are solved. While this computation takes about 1--2 days, the resulting 4D grid is readily interpolated. With the interpolation stored in computer memory, it is available for fast evaluation. MD simulations provide input on the effect of chemistry in the form of the function $q(t,\nu)$, which is different for scenarios 1 and 2. MD simulations are also used to find the relationship between \Tzz\ and the number of trapped HD$^+$ ions during secular excitation, which determines the shape of $f_\text{NL}$ (see Appendix~\ref{nlfunction}). This relationship (and therefore $f_\text{NL}$) depends on the different ion numbers used in scenarios a and b.}
\end{figure}

A schematic overview of the various steps involved in the construction of the fit function $S_\text{fit}$, as well as the role of the four scenarios, is presented in Fig.~\ref{fig:Sfitschematic}.

\subsection{\label{bg}Background gas reactions}
During the REMPD phase the number of HD$^+$ ions is not only reduced through photodissociation by the lasers. As described in Sec.~\ref{chemistry}, trapped ions can react with residual H$_2$ molecules in the vacuum setup. In order to correct the spectroscopic signal for these so-called background losses, the rates of reactions 2 and 3 in Table \ref{tab:chemreactions} are measured and included into the rate equations, Eq.~(\ref{eq:rateeq}), as an additional loss channel.

\begin{figure}
\centering
\includegraphics[trim=0cm 0cm 0cm 0cm, clip, width=6.5cm]{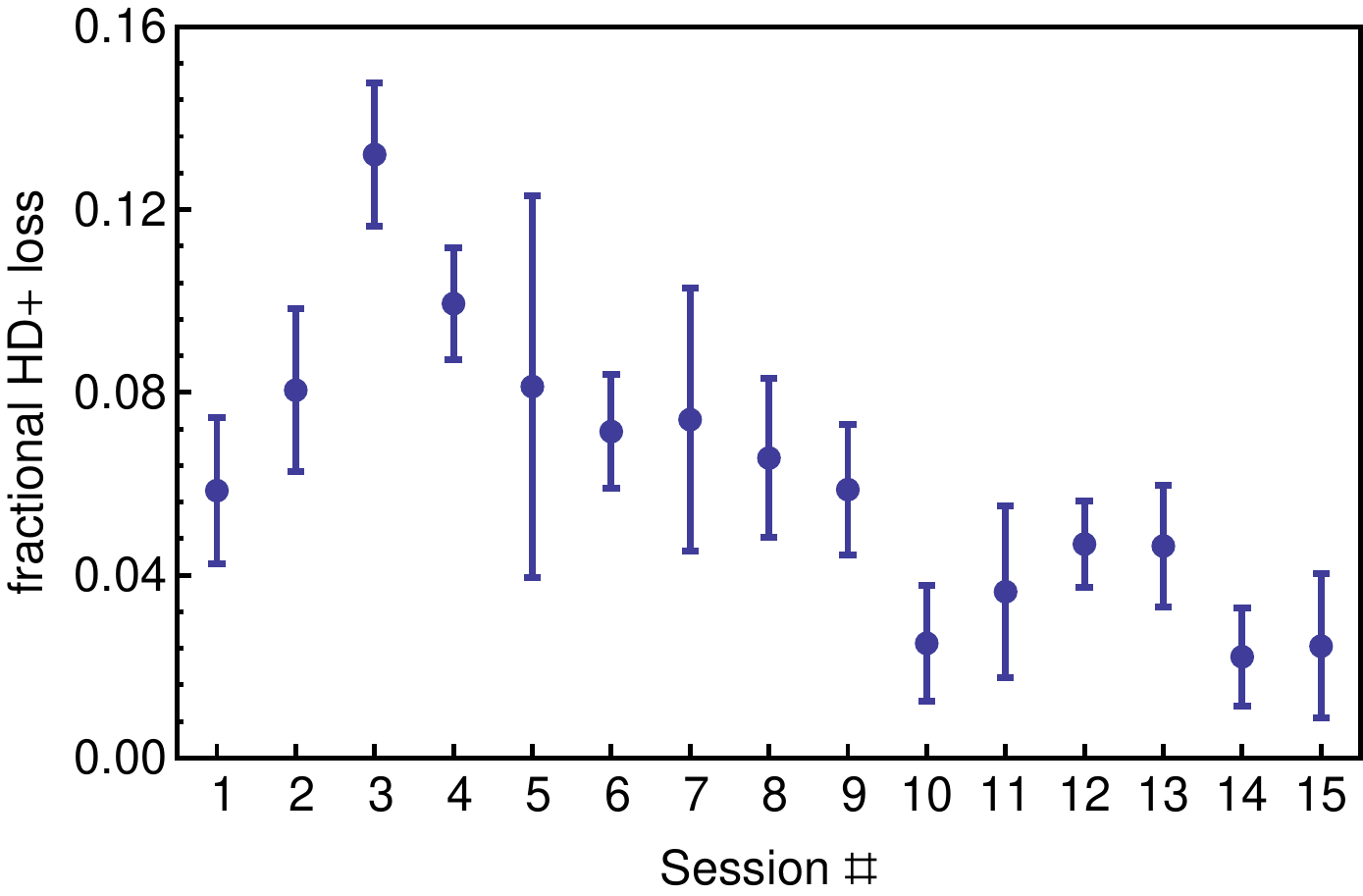}% Here is how to import EPS art
\caption{\label{Backgroundplot}Signals corresponding to background gas losses per measurement session. During some sessions the background pressure of the vacuum was higher, which results in a higher signal. The dots are the averages of a set of measurements, the error bars represent a $\pm$1$\sigma$ standard deviation.}
\end{figure}

The spectral data were acquired over the course of several months during 15 independent measurement sessions lasting several hours each (Fig.~\ref{Backgroundplot}). During this period the background pressure varied from session to session. For each session, the \HD\  background signal is obtained by using the measurement scheme depicted in Fig.~\ref{fig:ssREMPDss}, but with a shutter blocking the 782~nm laser, thus preventing REMPD. Typically, a few \HD\ ions react with H$_2$ which is detected as a small difference between the secular scan peak areas $A_{\text{bg,i}}$ and $A_{\text{bg,f}}$. During each measurement session, a series of $\sim 7$ background loss measurements is carried out three times, providing a data set of about 20 background measurement per session. From the average background loss signal per session a reaction rate $\gamma_{\text{bg}}$ is extracted from the relation
\begin{equation}\label{eq:gammadet}
1-e^{-\gamma_{\text{bg}} t}=\frac{N_{\text{bg,i}}-N_{\text{bg,f}}}{N_{\text{bg,i}}}= f^{-1}_{\text{NL}} \left( \Tz,\frac{A_{\text{bg,i}}-A_{\text{bg,f}}}{A_{\text{bg,i}}} \right)
\end{equation}
with $t=10$ s and $f^{-1}_{\text{NL}}$ mapping $(A_{\text{bg,i}}-A_{\text{bg,f}})/A_{\text{bg,i}}$ onto $(N_{\text{bg,i}}-N_{\text{bg,f}})/N_{\text{bg,i}}$ (see Appendix~\ref{nlfunction}).
%The estimated value of  $\Tz$ = 3 K is inserted in equation \ref{eq:gammadet} and after a few steps of iteration (see figure \ref{iteration}) a value of 3.5 K is obtained.
The values of $\gamma_{\text{bg}}$ are inserted into a modified set of rate equations, which include the process of background loss reactions:
\begin{equation}\label{eq:rateeqbg}
\frac{d \boldsymbol{\rho_{bg}}(t)}{d t}=\left( M_{\text{rempd}}\ + M_{\text{BBR}} + M_{\text{bg}} \right) \boldsymbol{\rho_{bg}}(t).
\end{equation}
Here, the vector $\boldsymbol{\rho_{bg}}$ equals $\boldsymbol{\rho}$ of Eq.~(\ref{eq:rateeq}) extended by two additional rows which describe the occurence of ions in the form of \HHD\ or \HHH. $M_{\text{bg}}$ is a diagonal matrix containing $\gamma_{\text{bg}}$ which describes the \HD\ losses. We take into account the fact that conversion of HD$^+$ to H$_3^+$ in reaction 3 in Table~\ref{tab:chemreactions} is not detected by the method of detection through secular excitation because \HD\ and \HHH\ have the same mass-to-charge ratio. This also means that the measured values of $\gamma_{bg}$ only represent the rates for reaction 2. As explained in Sec.~\ref{chemistry}, the reaction rate of 3 is a factor of three lower, which is taken into account as well.

We correct the data set of each session individually for the background signal following an iterative procedure. During the first step we insert an initial (coarse) estimate of the value of $\Tz$ in Eq.~(\ref{eq:gammadet}), and we simply subtract the signal predicted by the model without background losses (based on Eq.~(\ref{eq:rateeq})) from the signal prediction including background losses (based on Eq.~(\ref{eq:rateeqbg})). In this way an estimate of the background signal is obtained which is subsequently subtracted from the raw measurement data. Then the spectral fit function $S_\text{fit}$  (see Sec.~\ref{lsmodel}) is fitted to the corrected data points with free fit parameters $\Tz$, $T_{\text{HD}^+}$, $I_{\text{laser}}$ and $\nu_{0,\text{fit}}$. To find an improved estimate of the background signal, the thus found values of $\Tz$, $T_{\text{HD}^+}$, $I_{\text{laser}}$ are reinserted into Eqs.~(\ref{eq:rateeq}) and (\ref{eq:rateeqbg}), and the above procedure is repeated. After a few iterations, convergence is achieved.

The apparent loss of \HD\ ions is also influenced by reactions 4 and 5 in Table~\ref{tab:chemreactions}, which result in the production of additional \HHH\ during REMPD, leading to a reduction of the observed REMPD and background signals mentioned above. This effect cannot be easily assessed experimentally, and instead we estimate it for both scenarios a and b using the reaction rates of Table~\ref{tab:chemreactions}, leading to slightly modified values of $N_i$ and $N_f$. This translates to a small correction to $\epsilon$ in Eqs.~(\ref{eq:epsilon}) and (\ref{eq:gammadet}); see also Fig.~\ref{fig:Sfitschematic}). In addition, reactions with background gas also take place during the secular scans, thus influencing the determination of the initial and final numbers of particles with charge-to-mass ratio 1:3 itself. For example, the conversion of \HD\ to H$_2$D$^+$ (reaction 2) and to H$_3^+$ (reaction 3) have a small effect on the number of estimated \HD\ ions from the secular scan. Again, we can readily correct for these effects, knowing the values of all the relevant reaction rates, and noting that the reactions take place on a much longer timescale than the secular scan itself so that constant production rates can be assumed. Altogether, the corrections to $N_i$ and $N_f$ are at the level of a few percent, and are applied through Eqs.~(\ref{eq:epsilon}) and (\ref{eq:gammadet}) (see also Fig.~\ref{fig:Sfitschematic}).
\subsection{\label{dataproc}Spectrum, systematic effects and final result}
\begin{figure}
\centering
\includegraphics[width=\columnwidth]{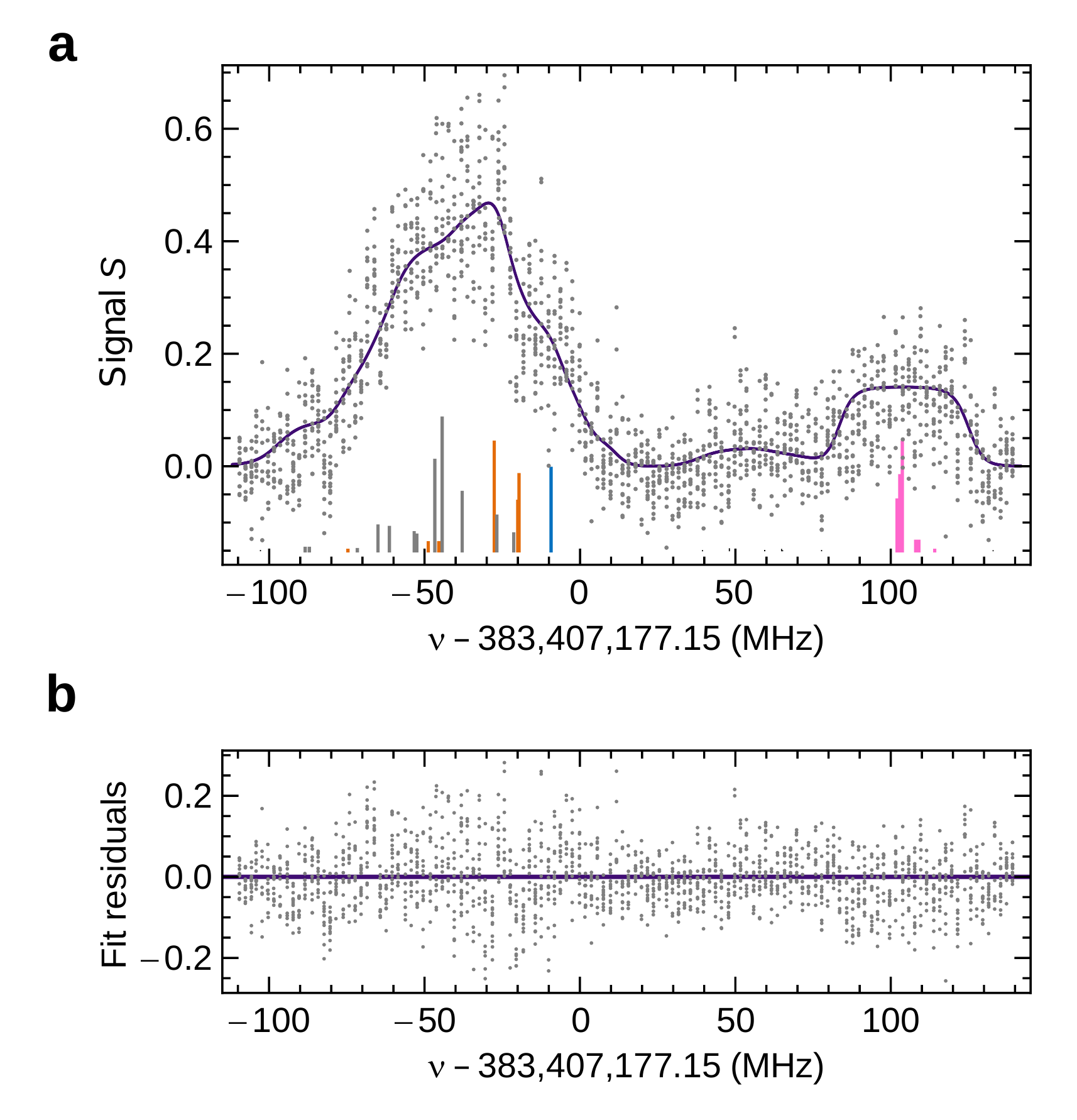}% Here is how to import EPS art
\caption{\label{Spectrum}(a) Measured spectrum (dots) of the \vvL\ transition and a least-squares fit (solid curve). All 1772 data points are plotted together with the fit function corresponding to scenario 1a. The fitted functions corresponding to the different scenarios are visually indistinguishable from the result shown here. The colored sticks represent the positions and linestrengths of the theoretical individual hyperfine components, and are plotted using the same color coding as used in Fig.~\ref{REMPDabc}(b). (b) Fit residuals.}
\end{figure}

The function $S_{\text{fit}}$ (Eq.~(\ref{Sfit2})) is fitted to the REMPD data set after correction for the background signal. Figure \ref{Spectrum} shows the REMPD data set and fit function for scenario 1a.
The noise in the spectrum has several origins. Firstly, the number of trapped ions is relatively small and varies from shot to shot. Secondly, the population in the various hyperfine states of the $L$=2 state varies from shot to shot, as expected for hyperfine states with a mean occupancy of order unity. The (stochastic) BBR interaction, which couples the states with $L=2$ with other rotational states, introduces additional random signal variations. Furthermore, the variation of the number of reactions of \HD\ with the background gas is in the order of a few per shot, which dominates the noise for low REMPD signals. Finally, part of the noise originates from random intensity variations due to spatial alignment variations of the 313~nm, 782~nm and 532~nm lasers.

\begin{figure}
\centering
\includegraphics[trim=0cm 0cm 0cm 0cm, clip, width=6.5cm]{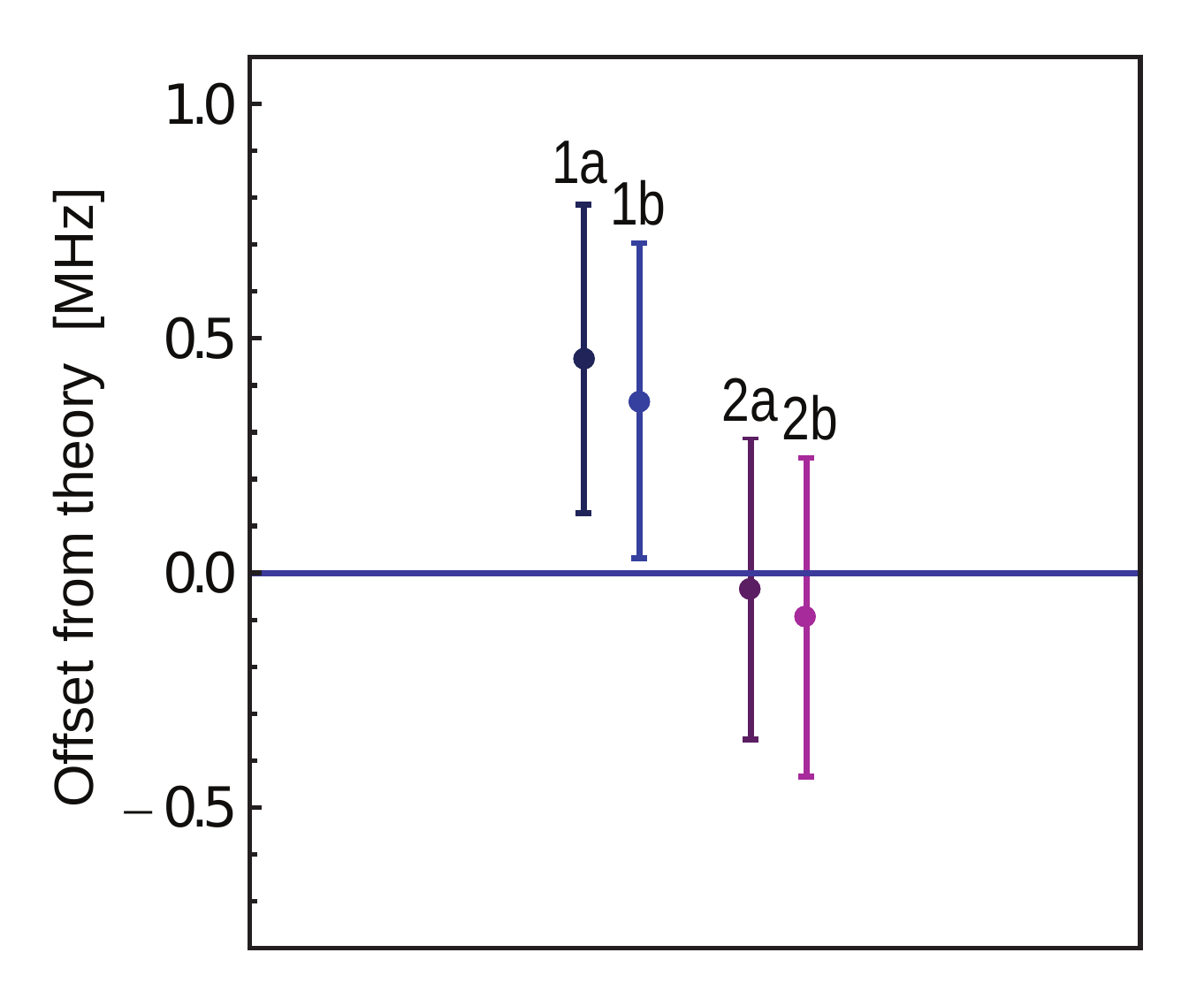}
\caption{\label{v0v8results4}Values of $\nu_{0,\text{fit}}$-$\nu_{\text{th}}$ found for the four different scenarios 1a, 1b, 2a and 2b. Error bars indicate the 1$\sigma$ fit uncertainty. All values are plotted with respect to the theoretical frequency (blue horizontal line)~\cite{Korobov2014b}.}
\end{figure}

 %We chose to use the most straightforward method to improve the signal-to-noise ratio (SNR) which is the increase of the number of data points (increase number of measurements). Other methods to improve the SNR are selection of the rotational states (rotational cooling) with lasers \cite{Shen2012}, cryogenically cooling the setup and laser power stabilization, but those methods require a lot of time and effort and therefore we chose not to use them. Loading only a few ions per shot, which allows counting of absolute and integer numbers of \HD\ losses, will also improve the SNR, but is obviously at the expense of time efficiency. \\
For each of the four scenarios (1a,1b,2a,2b) we obtain a particular set of fit parameters, which are listed in Table \ref{tab:fitpars}.
\begin{table*}%The best place to locate the table environment is directly after its first reference in text
\small
\caption{\label{tab:fitpars}%
The fit results of the free fit parameters per scenario. Scenarios 1 and 2 result in significantly different values of $\nu_{0,\text{fit}}$. Scenario a and b give give different results for $\Tz$. This  can be explained by the different amounts of trapped molecular species, which result in different \Be\ temperatures during a secular scan.
}
%%\begin{ruledtabular}
%\begin{tabular}{lllll}\hhline{=====}
\begin{tabular*}{1\textwidth}[t]{@{\extracolsep{\fill}}lllll}
\hline
\hline
& \multicolumn{3}{c}{\textrm{Scenario}}\\
\cline{2-5}
fit parameter& 1a & 1b & 2a & 2b\\
\hline
%%\colrule
$\nu_{0,\text{fit}}-\nu_{\text{th}}$(MHz) & 0.46(33)&0.37(34)&-0.03(32) &-0.12(33) \\
T$_{\textrm{HD}^+}$(mK)& 10.9(8) &11.0(8)  &10.6(8)  &10.35(80) \\
I$_{\textrm{laser}}$ ($\times$10$^7$ Wm$^{-2}$)& 0.99(14) &1.04(15) &0.95(14) &0.98(15) \\
$\Tz$(K)& 2.79(3) &3.85(5)&2.82(5)&3.84(5) \\
\hline \hline
\end{tabular*}
%%\end{ruledtabular}
\end{table*}
The correlation coefficients of the fit parameters are presented in Table~\ref{tab:CorrM} and
the values for $\nu_{0,\text{fit}}-\nu_{\text{th}}$ are graphically shown in Fig.~\ref{v0v8results4}.
The error bars represent the $\pm 1\sigma$ fit uncertainty which can be considered as the purely statistical precision of the spectroscopy measurement. We remark that the sensitivity to the chemistry processes (scenarios 1 and 2) is much stronger than the sensitivity to the numbers of \HD\ ions in the trap (scenarios a and b), and that the values $\nu_{0,\text{fit}}$ for 1a and 2a represent extreme upper and lower limits (with respect to the line shift due to chemistry) for the average number of \HD\ ions. We therefore chose to obtain our final result for $\nu_{0,\text{fit}}$ by taking the mean of these two values, while interpreting the mean single-fit error of (which is virtually the same for all four results) as the statistical uncertainty of the final result. We subsequently quantify the 'which-scenario' uncertainty as follows. For scenarios 1a and 1b (and similarly for 2a and 2b) the difference is due to a $1\sigma$ variation in the number of \HD\ ions. Therefore, we treat the frequency interval between the two values corresponding to scenario a and b as the corresponding $1\sigma$ interval, which amounts to 80~kHz when averaged over the two scenario's 1 and 2. To find the error corresponding to scenarios 1 and 2, we take the frequency interval between the values found for scenarios 1a and 2a (which are essentially extreme limits) and conservatively equate the interval to a 68~\% confidence interval. The interval thus corresponds to 2$\sigma$, with $\sigma=0.23$~MHz. The uncertainty of 0.5~mK in the temperature difference between top and wings of the spectrum (see Sec.~\ref{chemistry}) results in 28~kHz difference in $\nu_{0,\text{fit}}$, which is treated as a $1\sigma$ variation. The frequency shifts due to these systematic effects are listed together with their uncertainties in Table~\ref{tab:errorbudget}.

\subsubsection{Frequency uncertainty of the 782~nm laser}
\begin{figure}
 \centering
\includegraphics[trim=0cm 0cm 0cm 0cm, clip, width=8cm]{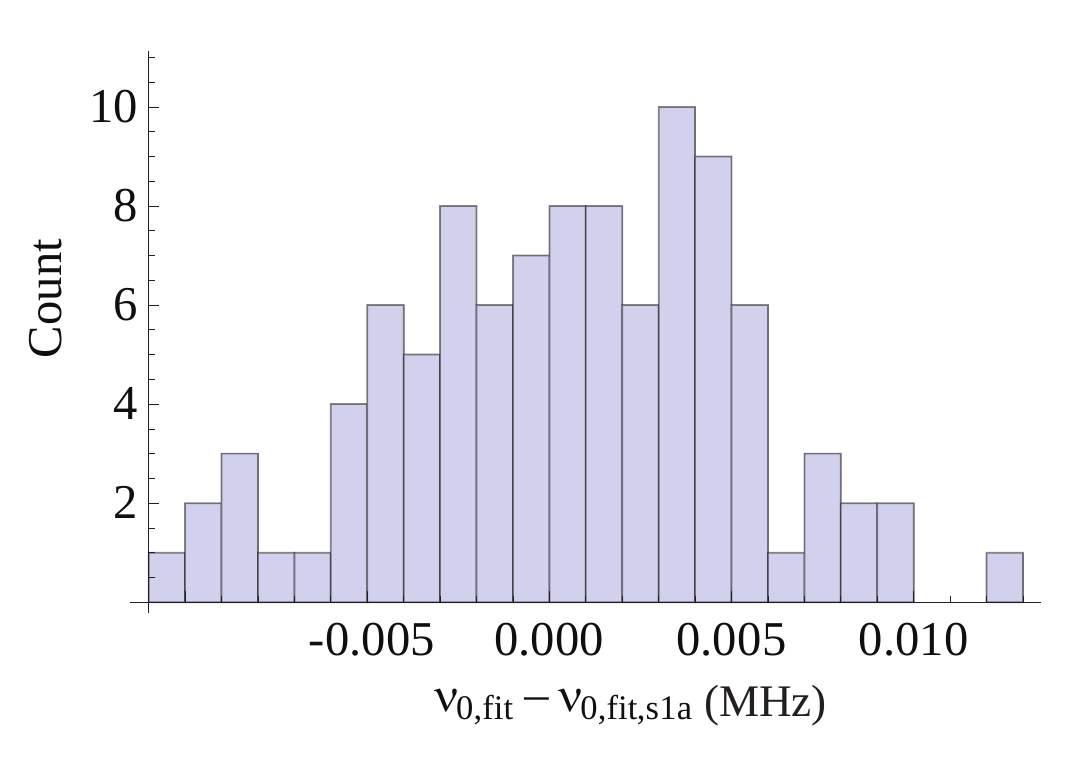}
\caption{\label{histogram1}Histogram of 100 fitted frequencies of the \vv\ transition, obtained from a Monte Carlo simulation involving 100 data sets with added random frequency noise (with the noise level corresponding to the measured noise level; see text). On the horizontal axis, zero corresponds to $\nu_{0,\text{fit}}$ obtained for scenario 1a. The mean offset and standard deviation are 0.5~kHz and 8~kHz, respectively.}
\end{figure}

The beat note of the frequency-locked 782~nm laser with the optical frequency comb is counted during REMPD. We use the beat-note frequencies to compute the Allan deviation, which is of the order of 0.1~MHz after 10~s averaging.  The uncertainty of the 782 nm laser frequency can be transferred to an uncertainty in the \vv\ fit result $\nu_{0,\text{fit}}$  by taking the Allan deviation as a measure of the standard deviation of a Gaussian noise distribution, describing the laser frequency offset from the set frequency during each REMPD cycle. We perform a Monte Carlo simulation in which  each of the 140 measurement frequencies is assigned a frequency offset, selected at random from the Gaussian distribution. Repeating this 100 times generates 100 different spectral data sets. Fitting $S_\text{fit}$ to each of the data sets gives 100 different values of $\nu_{0,\text{fit}}$. A histogram of the resulting distribution of $\nu_{0,\text{fit}}$ values is shown in Fig.~\ref{histogram1}. From the histogram we find a mean offset 0.5~kHz from the frequency value $\nu_{0,\text{fit}}$ found for scenario 1a, and a standard deviation of 8~kHz. We conclude that frequency noise introduces no significant bias, and we conservatively assume the uncertainty of the frequency measurement to be 0.01~MHz.
\begin{table}[b]
\caption{\label{tab:CorrM} The correlation coefficients for the fit of $S_\text{fit}$ (scenario 1a) to the data.
}
%%\begin{ruledtabular}
\begin{tabular*}{\columnwidth}{@{\extracolsep{\fill}}lllll}
% & \multicolumn{3}{c}{\textrm{Scenario}}\\
%\cline{2-5}
\hline
\hline
& $\nu_{0,\text{fit}}-\nu_{\text{th}}$ & T$_{\textrm{HD}^+}$& I$_{\textrm{laser}}$  & $\Tz$\\
%%\colrule
$\nu_{0,\text{fit}}-\nu_{\text{th}}$ &1 &-0.609&0.483 &0.049 \\
T$_{\textrm{HD}^+}$& -0.609 & 1 & -0.690&0.038\\
I$_{\textrm{laser}}$&0.483&-0.690 &1 &0.579 \\
$\Tz$& 0.049 &0.038 & 0.579&1 \\
\hline
\hline
\end{tabular*}
%%\end{ruledtabular}
\end{table}

The 782~nm laser has a Gaussian lineshape with a width of $\sim$0.5~MHz. The convolution of this lineshape with the (Gaussian) Doppler-broadened line (16~MHz) will give rise to another Gaussian lineshape. Since the linewidths add up quadratically, the increase in linewidth is smaller then the uncertainty of the linewidth due to the fit uncertainty of the temperatures, which is $\sim$~0.8~mK. Therefore, we consider the laser linewidth to be completely absorbed into the fitted temperature $T_{\text{HD}^+}$ with no significant effect on its value.

\subsubsection{Zeeman, Stark and other shifts}\label{Zeeman}
\begin{figure}
\centering
\includegraphics[trim=0cm 0cm 0cm 0cm, clip, width=7cm]{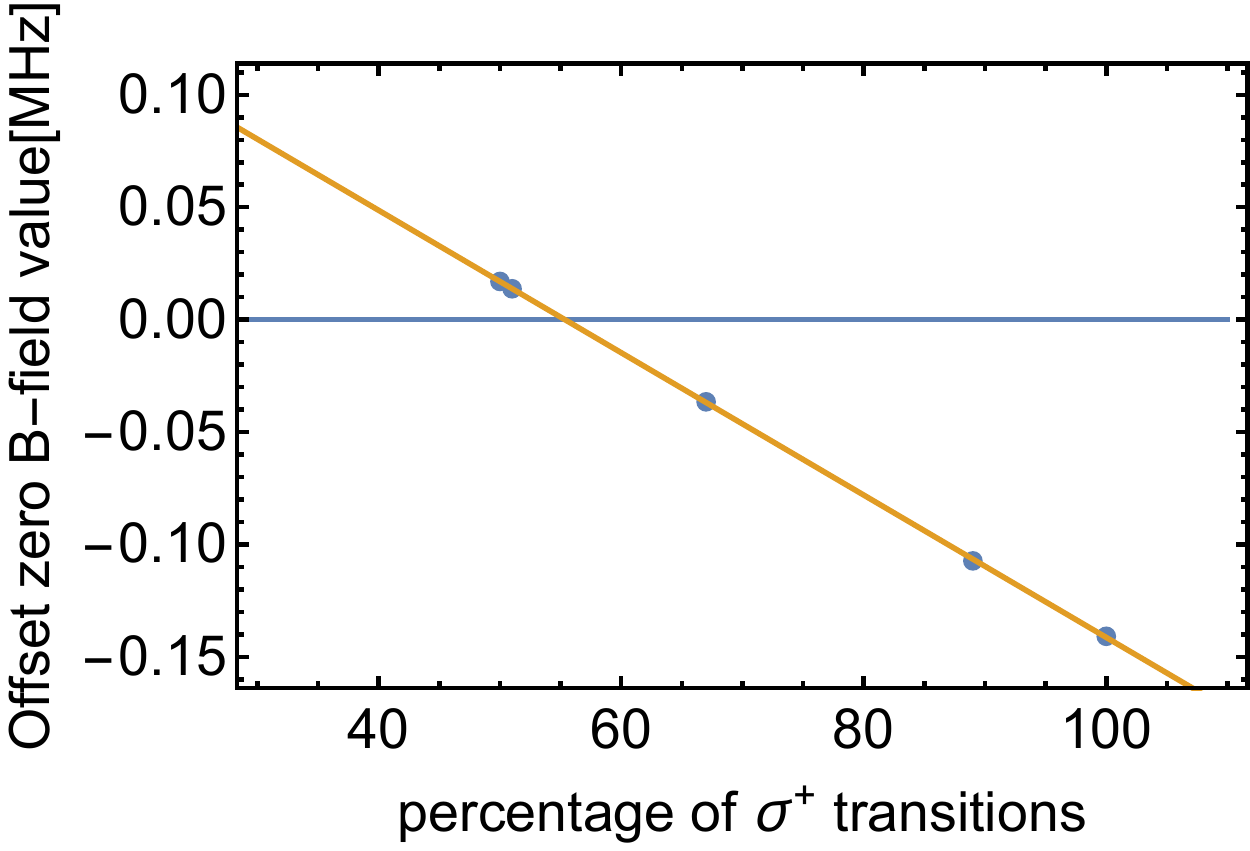}
\caption{\label{Zeemanplot}Zeeman shift of $\nu_{0,\text{fit}}$ found with a spectral fit function including the Zeeman effect due to a 0.19~mT magnetic field, for different ratios of $\sigma^+ / \sigma^-$ transitions. The horizontal line at 0~MHz corresponds to the value obtained using a fit function assuming zero magnetic field.}
\end{figure}

So far we have neglected the Zeeman splitting of the lines in the spectrum. Incorporating the Zeeman effect makes the hyperfine transition matrices very large and $\textsc{Mathematica}$ is only able to solve the rate equations [Eq.~(\ref{eq:rateeq})] effectively if lineshapes are not too complicated. We circumvent these issues as follows. First, we calculate the lineshifts and linestrenghts of the magnetic subcomponents of individual hyperfine lines. This is done by diagonalizing the sum of the hyperfine and Zeeman Hamiltonians (using the 0.19~mT field value used throughout the experiments), after which the eigenvectors and energy values are used to compute the stick spectrum of the \vvL\ transition. This procedure is similar to that followed in Refs.~\cite{Koelemeij2007a,Bakalov2011}, and can readily be done for the case of $\sigma^+$, $\sigma^-$ and $\pi$ transitions, and superpositions thereof. As the Zeeman splitting is small compared to the Doppler width, the magnetic subcomponents belonging to the same hyperfine component overlap well within the profile of the lineshape function $D_z$, forming a new composite (and Zeeman-shifted) lineshape function $D'_z$. This new lineshape function is subsequently used in Eq.~(\ref{eqMrempd}). For simplicity, we do not implement effects of micromotion and chemistry in this analysis, and we compare $\nu_{0,\text{fit}}$ fit results based on versions with $B=0.19$ mT with a version with zero $B$ field. It is important to note that the linear polarization of the 782~nm laser is practically perpendicular to the $B$ field, so that during the \vv\ excitation only $\sigma^+$ and $\sigma^-$ transitions are driven. Due to polarization imperfections, caused by the polarization optics and the slightly birefringent viewports of the vacuum chamber, the two circularly polarized components are estimated to have a maximum possible intensity imbalance of 2\%. Figure~\ref{Zeemanplot} shows the offset between the $\nu_{0,\text{fit}}$ fit results, obtained with the above model for several $\sigma^+$/$\sigma^-$ intensity ratios, and the fit result we found previously assuming zero magnetic field. For the 0.19~mT field used in our experiment, a small shift of -0.017(3)~MHz is obtained, with the uncertainty due to the possible maximum polarization imbalance.

\begin{table*}[t]%The best place to locate the table environment is directly after its first reference in text
\caption{\label{tab:errorbudget}%
Systematic shifts and uncertainty budget.
}
%%\begin{ruledtabular}
\begin{tabular*}{\textwidth}{@{\extracolsep{\fill}}llll}
\hline
\hline
\textrm{Origin}&\textrm{Shift}&\multicolumn{2}{c}{\textrm{Uncertainty}}\\
\cline{3-4}
&\textrm{(MHz)}&\textrm{(MHz)}&\textrm{(ppb)}\\
%%%\colrule
\hline
Resolution (statistical fit error) & 0 & 0.33&0.85 \\
Uncertainty $q$ value & -0.25$^\text{a}$ & 0.23&0.61 \\
Uncertainty $N_{\text{HD}^+}$& 0 & 0.080&0.21 \\
Ignoring populations $L$=6 &0 & 0.032 & 0.083 \\
$T_{\textrm{HD}^+}$ variation in spectrum & 0 & 0.028& 0.072 \\
Doppler effect due to micromotion&-0.055$^\text{a}$ & 0.020 & 0.052 \\
Frequency measurement &0 & 0.010 & 0.026 \\
BBR temperature &0& 0.005 & 0.013 \\
Zeeman effect &-0.0169 & 0.003 & 0.008 \\
Stark effect &-0.0013& 0.0001 & 0.0004 \\
Electric-quadrupole shift &0$^\text{b}$ & 0.0001 &0.0003 \\
2$^{\textrm{nd}}$ order Doppler effect&0$^\text{b}$ & 0.000005 & 0.00001 \\
Uncertainty $E_4$&0$^\text{b}$& 0.000001 & 0.000003\\
\hline
Total&-0.0182& 0.41 & 1.1 \\
\hline
\hline
\end{tabular*}
\begin{tabular*}{1\textwidth}[t]{@{\extracolsep{\fill}}p{1\textwidth}}
{\begin{footnotesize}$^\text{a}$This is a shift with respect to a scenario with a zero effect of $q$
or micromotion, and serves to illustrate the size of the effect. The shift itself, however, is absorbed in the value of $\nu_{0,\text{fit}}$.
 \end{footnotesize}}  \\
{\begin{footnotesize}$^\text{b}$The value of these shifts is actually nonzero but negligibly
small, and therefore ignored here. \end{footnotesize}}
\end{tabular*}
%%\end{ruledtabular}
\end{table*}

The ac Stark shifts due to the 782~nm, 532~nm and 313~nm lasers are $-$869~Hz, $-$452~Hz and 8~Hz respectively.  These values represent the shift of the center of gravity of the spectral line and can therefore be considered as weighted means of all the shifts of the single hyperfine components. The calculation of the Stark shifts is shortly explained in Appendix~\ref{Starkshift}. Stark shifts due to the BBR and trap rf field are calculated in \cite{Koelemeij2011} and are smaller than 1~Hz. Together, this gives us a total Stark shift of 1.3(1)~kHz. The uncertainty in this value stems almost exclusively from the accuracy to which the laser beam intensities are known.

A conservative upper limit of 100~Hz to the electric-quadrupole shift for the \vvL\ transition is obtained from~\cite{Bakalov2014}. The second-order Doppler effect is dominated by average micromotion velocity of the \HD\ ions, and is found to be less than 5~Hz. The values of the aforementioned shifts and their uncertainties are listed in Table~\ref{tab:errorbudget}.

If we compare $\nu_{0,\text{fit}}$ from scenario 1a based on the model containing rotational states $L\le$5 with an extended version containing $L\le$6 states we find a shift of $\nu_{0,\text{fit}}$ of 28~kHz, which we treat as the uncertainty due to the neglect of population in $L=6$.  The rate-equation model furthermore includes the BBR temperature, which we estimate to be 300~K with an uncertainty of about 5~K, caused by day-to-day variations of the temperature in the laboratory, and by a possibly elevated temperature of the trap electrodes due to rf current dissipation. If we compare the $\nu_{0,\text{fit}}$ values after inserting BBR temperatures of 300~K and 305~K, we obtain a difference of 5~kHz, which we include in the uncertainty budget. Micromotion also causes a shift: by comparing the $\nu_{0,\text{fit}}$ value from scenario 1a that includes micromotion (amplitude 11(4)~nm) with a the result of a version with zero micromotion, a shift of  0.055(20)~MHz is obtained. Finally, we investigated the effect due to the uncertainty of the spin coefficient $E_4$ (see Eq.~(\ref{eq:hyperfine1})), which is estimated to be 50~kHz~\cite{Bakalov2006}. Comparing fits with $E_4$ values differing by 50~kHz we find that this has a negligibly small effect of $\sim$1~Hz on the result for $\nu_0$.

\subsubsection{Frequency of the \vvL\ transition} \label{nu0results}
We take the average of the fit values obtained from scenarios 1a and 2a, corrected by the systematic shifts described above, to find the value of the \vvL\ transition frequency, $\nu_0= 383,407,177.38$~MHz. Table~\ref{tab:errorbudget} shows the error budget, with a total uncertainty of 0.41~MHz that corresponds to 1.1~ppb. This result differs 0.21~MHz (0.6~ppb) from the more accurate theoretical value,  $\nu_{\text{th}}$= 383,407,177.150(15)~MHz~\cite{Korobov2014b}. The two main uncertainty contributions are the statistical fit error of 0.33~MHz and the uncertainty in the $q$-factor scenario of 0.23~MHz.

\section{Conclusion, implications and outlook}
\label{conclusion}
We have measured the \vvL\ transition in the \HD\ molecule with 0.85~p.p.b (0.33~MHz) precision, which is the first sub-p.p.b. resolution achieved in molecular spectroscopy. A thorough analysis of systematic effects points out that the total uncertainty is 1.1~p.p.b., and the result ($\nu_0= 383,407,177.38(41)$~MHz) differs by only 0.6~p.p.b. from the theoretically predicted value ($\nu_{\text{th}}=383,407,177.150(15)$~MHz). A large contribution to the systematic uncertainty is the effect of chemical reactions in the Coulomb crystal, of which the 1$\sigma$ uncertainty is 0.61~p.p.b (0.23~MHz). This effect, which had not been recognized before, causes a nonthermal velocity distribution that can be approximated by a $q$-Gaussian function, and which significantly influences the accuracy of laser spectroscopy of composite lineshapes in the presence of strong saturation and depletion of the HD$^+$ sample. This is a situation regularly encountered in laser spectroscopy of Doppler broadened transitions in finite samples of trapped molecular ions.

The agreement between experimental and theoretical data has several implications. First, it enables a test of molecular theory at the 1-p.p.b. level, and a test of molecular QED at the level of $2.7\times10^{-4}$, which are the most stringent tests performed so far. Second, it allows us to put new bounds on the existence of hypothetical fifth forces, and put new limits on the compactification radius of higher dimensions, as described in detail in~\cite{Salumbides2013,Salumbides2015,Biesheuvel2015}. Third, the result presented here can be used to obtain a new value the proton-to-electron mass ratio with a precision of 2.9~p.p.b.~\cite{Biesheuvel2015}, for the first time from molecular spectroscopy as proposed already four decades ago~\cite{Wing1976}.

Our analysis clearly demonstrates that the first-order Doppler effect is responsible for the largest contribution to the uncertainty. In fact, removing the first-order Doppler effect would render the frequency measurement as the largest source of error (0.026~p.p.b), thereby immediately improving the uncertainty by about a factor of 40. Such --and even larger-- improvements are possible using two-photon spectroscopy. For example, in Refs.~\cite{Tran2013,Karr2016} an experiment was proposed in which the $(v,L)=(0,3)\rightarrow(9,3)$ line in \HD\ is addressed through a two-photon transition with nearly degenerate photons. Using counter-propagating laser beams with a narrow linewidth, the Lamb-Dicke regime may be reached in the present apparatus, such that first-order Doppler broadening is entirely eliminated, while all other systematic effects could be controlled below the $1\times 10^{-13}$ uncertainty level. Such spectroscopy would allow more stringent tests of molecular theory and QED, tighter bounds on new physics at the {{\AA}ngstr\"om} scale, a competitive determination of the proton-electron mass ratio, and even contribute to the determination of several other fundamental constants including the Rydberg constant, the deuteron-electron mass ratio, and the proton and deuteron charge radii~\cite{Karr2016}.
\section*{Acknowledgements}
We are indebted to J. Bouma, T. Pinkert and R. Kortekaas for technical assistance, and to V. Korobov, E. Hudson and R. Gerritsma for fruitful discussions. This research was funded through the Netherlands Foundation for Fundamental Research on Matter (FOM), the COST action MP1001 IOTA, and the Dutch-French bilateral Van Gogh Programme. J.C.J.K. thanks the Netherlands Organisation for Scientific Research (NWO) and the Netherlands Technology Foundation (STW) for support. SURFsara (www.surfsara.nl) is acknowledged for the support in using the Lisa Compute Cluster for MD simulations.

\section{Appendices}

\appendix
\section{Molecular Dynamics simulations}\label{MDsim}
MD simulations are implemented in $\textsc{Fortran}$ code in order to realistically describe the dynamics of trapped and laser-cooled ions in the presence of the time-de\-pen\-dent trapping field, 313~nm photon scattering by the \Be\ ions, and fast ionic products from chemical reactions. A cycle of one time step starts by computing the sum of the forces acting on each ion, which consists of the Coulomb force, $\vec{F}_C$, the time-dependent force from the trapping field field, $F_\text{trap}$, and an optional rf electric field, $F_\text{SS}$, which drives the secular motion:
\begin{equation}\label{Ftot}
 \vec F_{\text{tot}}=\sum \mathbf{F}=\mathbf{F}_{\text{C}}+\mathbf{F}_{\text{trap}}+\mathbf{F}_{\text{SS}}.
\end{equation}
The radial and axial part of $\mathbf{F}_{\text{trap}}$ are given by
\begin{equation}
F_{\text{trap},x,y}=-\frac{Q V_0}{R^2}(x\hat{x}-y\hat{y})\cos(\Omega t)+ \frac{1}{2} Q \omega_z^2(x\hat{x}+y\hat{y})
\end{equation}
and
\begin{equation}
F_{\text{trap},z}=a z + b z^3 + c z^5,
\end{equation}
where $\omega_z$ is the secular angular frequency in the $z$-di\-rec\-tion. The constants $a$, $b$ and $c$ depend on the trap geometry, which are determined through a finite-element analysis performed with the software package \textsc{SIMION}. The forces exerted on each ion are calculated, and trajectories are obtained using the Velocity Verlet method~\cite{Verlet1967}.
%and Velocity Verlet integration~\cite{Verlet1967} provides the new velocities $\mathbf{v}_{i+1}$ and positions $\mathbf{x}_{i+1}$ at the $i^{\text{th}}$ time step $\Delta \tau$:
%\begin{eqnarray}
%\mathbf{v}_{i+1} &=& \mathbf{v}_{i} + (\mathbf{F}_{\text{tot},i}/m_j) \Delta \tau \nonumber \\
%\mathbf{x}_{i+1} &=& \mathbf{x}_{i} + \mathbf{v}_{i+1} \Delta \tau
%\end{eqnarray}
%where $m_{\text{j}}$ is the mass of the $j^\text{th}$ ion. The new positions, $\mathbf{x}_{i+1}$, are inserted %into Eq.~(\ref{Ftot}), after which the next cycle starts.
Doppler-cooling is included at the level of single-photon scattering. Photon momentum kicks are simulated as velocity changes where absorption only takes place in the laser direction. In order to include ion motional heating which occurs in the trap, we implemented additional stochastic velocity kicks with a size of the recoil momentum of a single 313~nm photon with random directions. If an average kick rate of 75~MHz is used, ion temperatures of around 10~mK are obtained.

The processes of elastic and inelastic neutral-ion collisions are simulated as velocity kicks in random directions. For example, simulating reaction 7 in Table~\ref{tab:chemreactions}, a \Be\ ion is substituted with a BeH$^+$ ion at 10~mK, after which its speed is modified so as to give it 0.25~eV of kinetic energy.

Simulation of particles which in some cases have high velocities requires the use of a variable time step size, $\Delta \tau$. If the proper step size is not observed, two particles with a high velocity difference at close distance could 'skip' each other within one time step instead of colliding. The default step size is $\Delta \tau = 0.2$~ns. However, if for any of the trapped particles the condition $ v_j \Delta \tau > 10 \times \text{min}\{\Delta x_{jk}\}$ is met, where $\text{min}\{\Delta x_{jk}\}$  is the distance between particle $j$ with velocity $v_j$ and the nearest particle $k$, the time step $\Delta \tau$ is reduced by a factor of 10. Likewise, the step size is increased if the colliding particles separate again and $ v_j \Delta \tau < 100 \times \text{min}\{\Delta x_{jk}\}$.

To simulate EMCCD images, we made use of a simpler MD implementation, which treats the motion of the ions in the pseudopotential approximation and which does not include high-energy ions. This allows for a larger integration time step (10~ns) and, thus, faster MD simulations.

\section{Derivation of the non-linear fluorescence function}\label{nlfunction}
The relative HD$^+$ loss during REMPD, $\epsilon$, is related to the spectroscopic signal $S$ through the non-linear function $f_{\textrm{NL}}$, which we derive here. The fluorescence yield during a secular scan depends on the Be$^+$ temperature, $T$, and is described by the scattering rate formula integrated over a Maxwell-Bolzmann velocity distribution. Neglecting micromotion effects, we take the scattering rate $R^{\text{MB}}=R^{\text{MB}}(T,\Delta,I/I_\text{sat})$ defined in Eq.~(\ref{eq:MMR}).
%%\begin{widetext}
%\begin{equation}
%\label{eq:NLRMB}
%\begin{split}
%R^{\text{MB}}(T) = & \frac{\Gamma}{2}\sqrt{\frac{m_{\text{Be}}}{2 \pi k_B %T}}\int\frac{I/I_\text{sat}}{I/I_\text{sat}+1+(2(\Delta-\textbf{k} v_k)/ \Gamma)^2}\\
%& \times \text{exp}\left(-\frac{m_{\text{Be}}v^2}{2k_B T}\right)dv_k,
%\end{split}
%\end{equation}
%%\end{widetext}
%where \textbf{v} along \textbf{k}, i.e. $v_k \equiv \textbf{k} \cdot \textbf{v}/ |\textbf{k}|$  is integrated over the distribution of \Be\ velocities.
During a secular scan in the experiment we use the values $\Delta$ = 2$\pi \times -$300~MHz and $I/I_\text{sat}=67$, and in what follows we drop these variables from the function argument of $R^{MB}$. While performing the secular scan, the temperature $T$ varies, which leads to a fluorescence peak as described by Eq.~(\ref{eq:MMR}). The spectroscopic signal $S$ is the relative difference between the areas under the fluorescence peaks (see Eq. (\ref{S})). We may rewrite the area, $A$, as
\begin{equation}\label{AppBA}
%A=C \left( \frac{R^{MB}(\bar{T})}{\Delta t}- \frac{R^{MB}(\bar{T}_{\text{bl}})}{{\Delta t}}\right),
A=C \left[ R^{\text{MB}}(\bar{T})\Delta t- R^{\text{MB}}(\bar{T}_\text{bl})\Delta t\right],
\end{equation}
where $C$ is a constant taking into account the collection and quantum efficiencies of the PMT or EMCCD imaging system, $\Delta t$ denotes the duration of the secular scan (10 s), and $\bar{T}$ stands for the `effective' value of $T$ during a secular scan. The effective temperature $\bar{T}$ is defined through Eq.~(\ref{AppBA}), and can be interpreted as follows. If we would fix the secular excitation field frequency and amplitude at a certain value during a scan, the \Be\ temperature and the fluorescence level would remain constant. $\bar{T}$ is the constant temperature that leads to the same area under the fluorescence trace as during a true secular scan. Likewise, $\bar{T}_{\text{bl}}$ stands for the effective baseline temperature during $\Delta t$ (corresponding to the fluorescence level that results if no \HD\ ions are present). In the experiment, the baseline may have a small slope due to the wing of the secular resonance of particles with mass 4 and 5~amu (see, for example, Fig.~\ref{SSm4m5ref}). This slope is detected and removed by the \textsc{Mathematica} code we use to analyze the PMT signal traces.

Both in the experiment and in the MD simulations described below, we observe the area $A$ under experimental or simulated fluorescence traces, to which we subsequently can assign an effective temperature $\bar{T}$ through Eq.~(\ref{AppBA}). We emphasize that no attempt is made to derive $\bar{T}$ directly from, for example, the simulated velocities of \Be\ ions. Also note that in practice we only use Eq.~(\ref{AppBA}) to assign effective temperatures to simulated fluorescence traces, as in this case the constant $C$ is known ($C=1$).

Inserting Eq.~(\ref{AppBA}) into Eq.~(\ref{S}), we obtain
\begin{equation}\label{AppBS}
S=\frac{R^{\text{MB}}(\bar{T}_i)-R^{\text{MB}}(\bar{T}_f)}{R^{\text{MB}}(\bar{T}_i)- R^{\text{MB}}(\bar{T}_{\text{bl}})},
\end{equation}
which is a relationship between the spectroscopic signal $S$ and the effective ion temperatures during the initial and final secular scan, $\bar{T}_i$ and $\bar{T}_f$ respectively.

\begin{figure}
 \centering
\includegraphics[width=6cm]{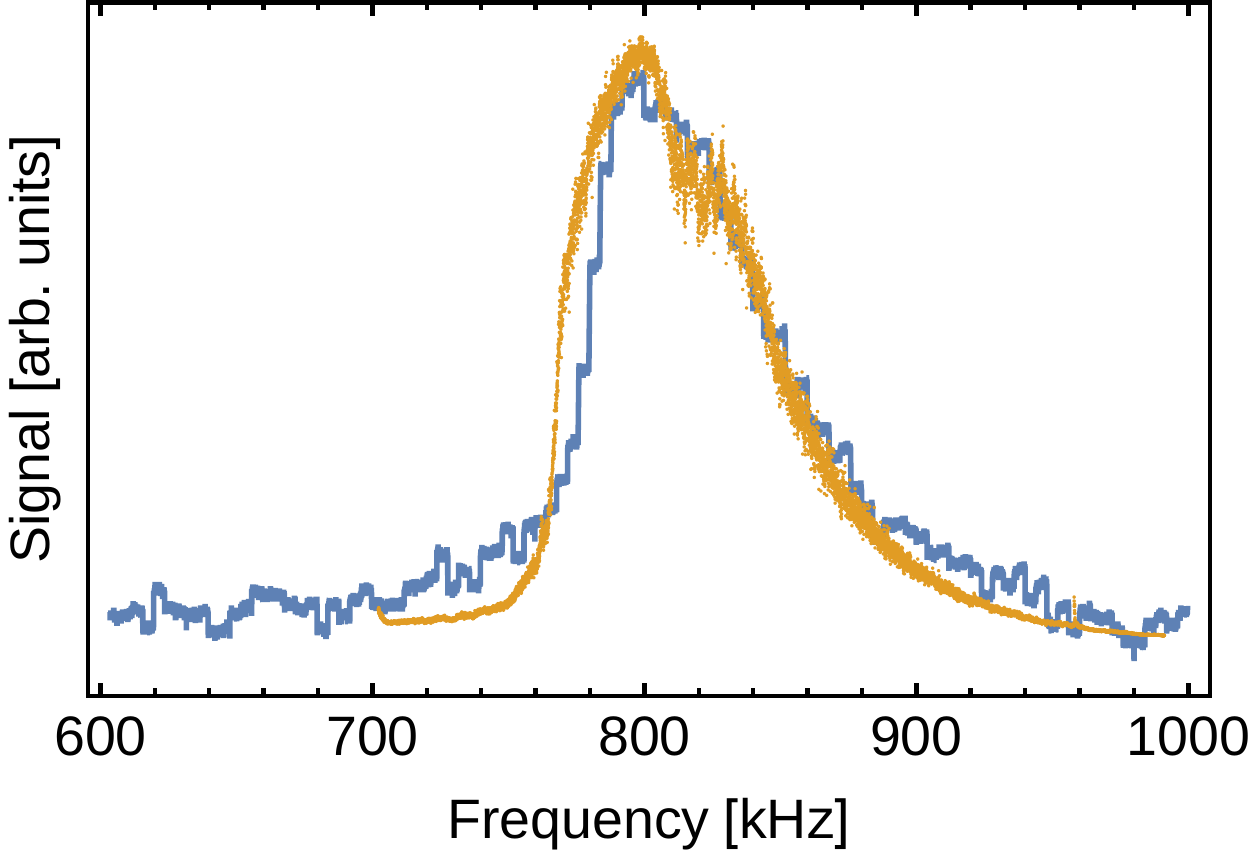}
\caption{\label{simSSpeak}A simulated (yellow) and a real (blue) secular scan peak plotted on the same frequency axis. The signals are scaled vertically to achieve matching peak heights.}
\end{figure}

The relationship between the number of trapped \HD\ molecules and $\bar{T}$ can be obtained from MD simulations. A Coulomb crystal is simulated containing 750 trapped Doppler-cooled Be$^+$ ions and with HD$^+$ numbers varying from 0 to 100. This is done once using a number of additional \HHD\ and \HDD\ ions equal to that of scenario a, and once using the numbers of \HHD\ and \HDD\ of scenario b (see Sec.~\ref{absN}). The simulated secular scans over the HD$^+$ secular resonance frequency produce fluorescence peaks which agree qualitatively with those obtained in the laboratory as shown in Fig.~\ref{simSSpeak}.\\
In the experiment, secular scans are acquired over a time span of 10~s. The dynamics due to the time-varying frequency of the ac electric field take place at a time scale much longer than the time scale of fluorescence dynamics during laser cooling, which takes place at time scales of the order of 10~$\mu$s~\cite{Wesenberg2007}, and also longer than typical ion oscillation periods and the frequency of the ac electric field itself (1--20~$\mu s$). However, due to limited computational resources, the simulated duration of a secular scan is approximately 100~ms, which implies that the simulated dynamics take place at a considerably faster rate than in the experiment, typically on the scale of milliseconds. However, this still is much longer than the time scale of fluorescence dynamics and the motional dynamics. Therefore, we assume that the simulated secular scan peaks provide a reliable model of the experimentally observed secular scans.

\begin{figure}
\centering
\includegraphics[width=7cm]{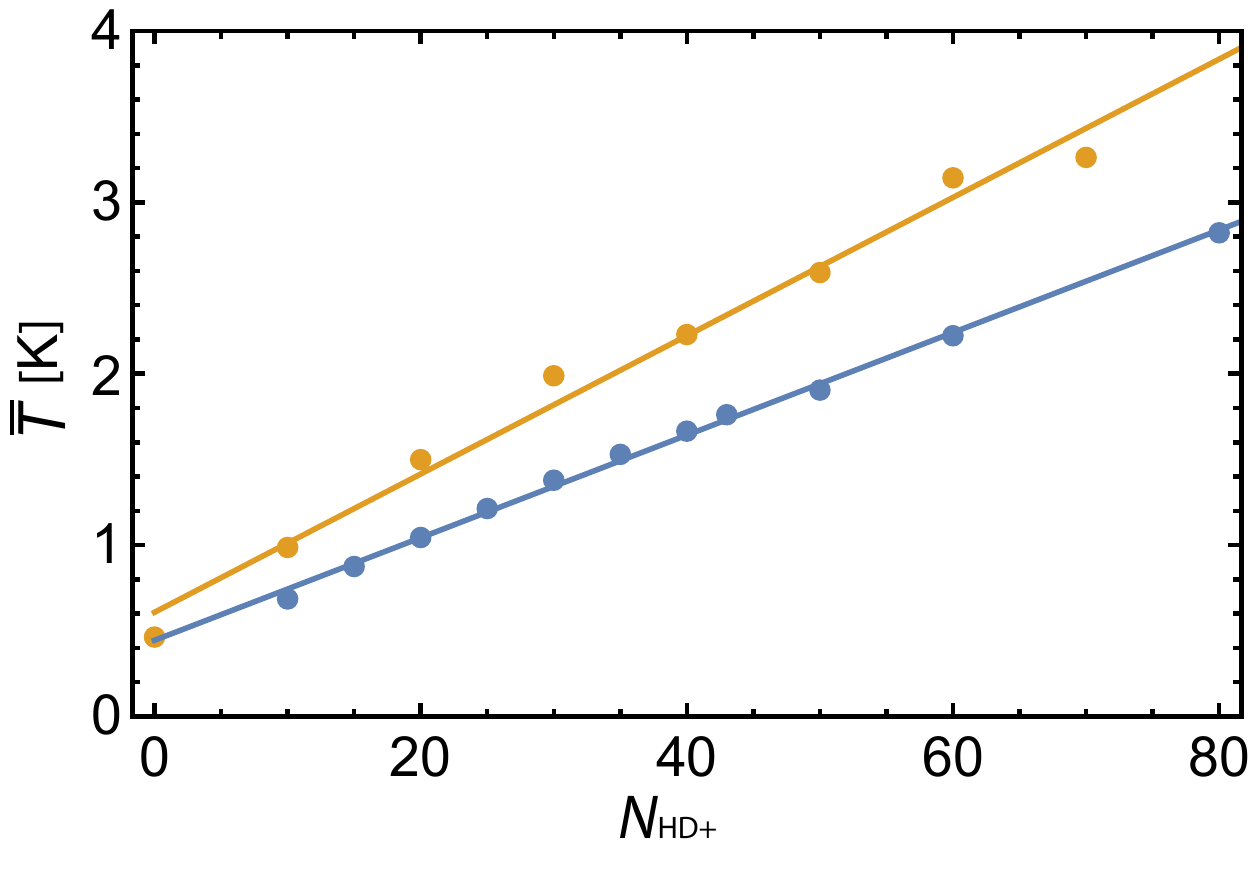}
\caption{\label{Tepsilon}Results from MD simulations showing the effective \Be\ ion temperature during a secular scan $\bar{T}$ versus the number of \HD\ ions $ N_{\text{HD}^+}$, assuming numbers of \HHD\ and \HDD\ ions as in scenario a (blue dots) and scenario b (yellow dots). The blue and yellow lines represent least-squares fits, revealing a linear relationship between $\bar{T}$  and $ N_{\text{HD}^+}$.}
\end{figure}

\begin{figure}
\centering
\includegraphics[width=7cm]{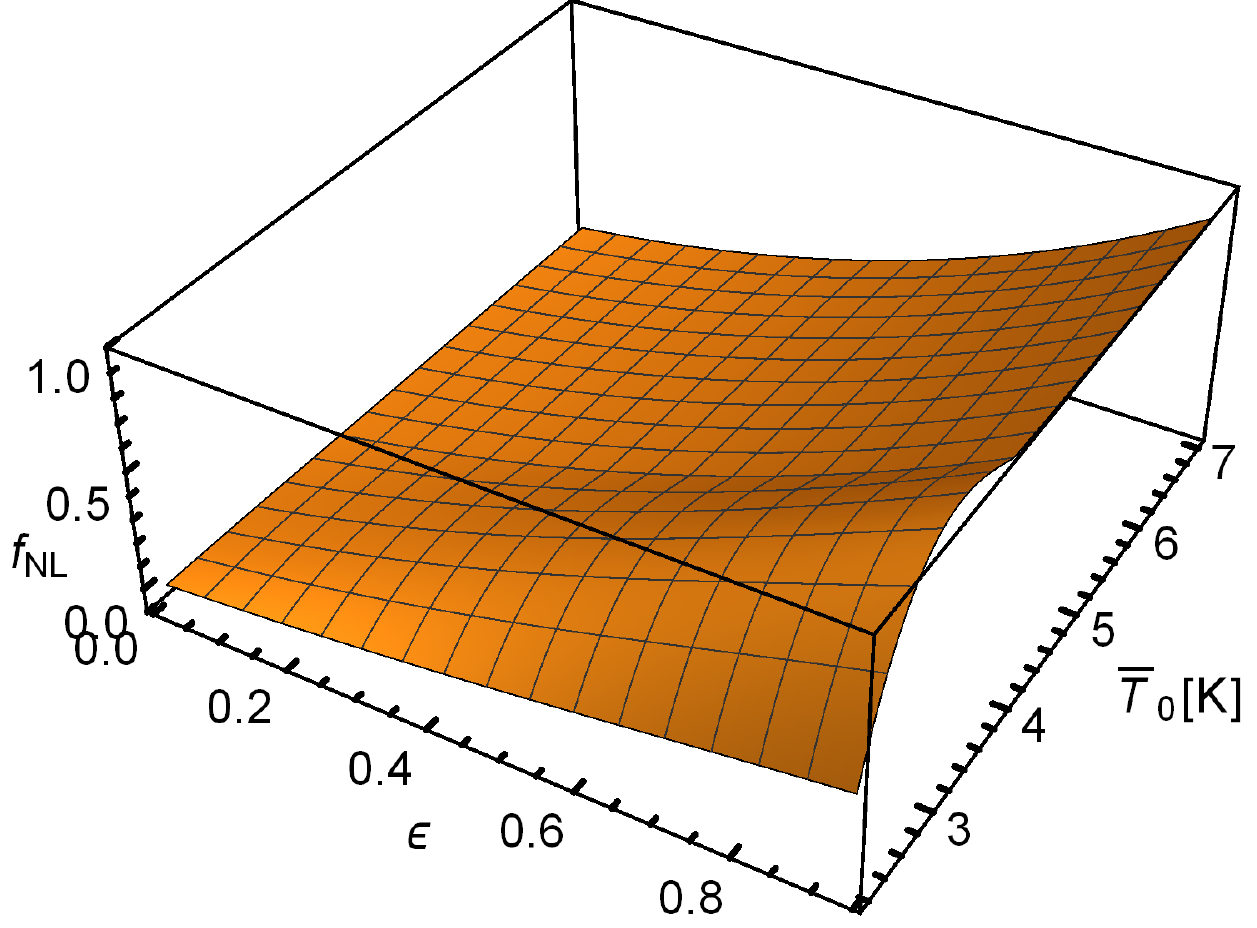}
\caption{\label{D3plotNLip15mVm4m5}
3D plot of the function $f_{\text{NL}}(\Tz,\epsilon)$ (here plotted for scenario a) which connects the raw measurement signal $S$ to the actual fractional loss of HD$^+$, $\epsilon$.}
\end{figure}

The MD simulations reveal a linear relationship between the number of trapped HD$^+$ ions and $\bar{T}$. This agrees with the intuitive picture of \Be\ ions with frictionally damped motion (because of the laser cooling), whose temperature rise during secular excitation is directly proportional to the number of \HD\ ions. Figure~\ref{Tepsilon} shows the ($N_{\text{HD}^+},\bar{T}$) relationship for the two scenarios a and b. Having established  that $\bar{T}$ is a linear measure of the number of trapped \HD\ molecules, we now combine the relations $\epsilon=(N_i-N_f)/N_i$, $\bar{T}_i=c_1 N_i+c_2$ and  $\bar{T}_f=c_1 N_f+c_2$, where $c_1$ and $c_2$ are constants derived from MD simulations (Fig.~\ref{Tepsilon}), to obtain
\begin{equation}\label{TfTi}
\bar{T}_f(\epsilon)=\bar{T}_i(1-\epsilon) + c_2 \epsilon.
\end{equation}
$\bar{T}_i$ can also be defined as the effective temperature with zero \HD\ loss ($\bar{T}_i=\bar{T}_f(\epsilon=0)\equiv \Tz$), while the term $c_2$ can be considered as the effective baseline temperature ($c_2=\bar{T}_f(\epsilon$=1)$= \bar{T}_{\text{bl}}$). Inserting Eq.~(\ref{TfTi}) into Eq.~(\ref{AppBS}) results in the nonlinear function
%\begin{widetext}
\begin{equation}\label{AppfNL}
f_{\text{NL}}(\Tz,\epsilon)\equiv \frac{R^{\text{MB}}(\Tz)-R^{\text{MB}}((\bar{T}_{\text{bl}}-\Tz)\epsilon +\Tz)}{R^{\text{MB}}(\Tz)- R^{\text{MB}}(\bar{T}_{\text{bl}})},
\end{equation}
%\end{widetext}
which is plotted for scenario a in Fig.~\ref{D3plotNLip15mVm4m5}. Note that the nonlinear dependence on $\epsilon$ originates from the nonlinear dependence of $R^\text{MB}$ on $T$ in Eq.~(\ref{eq:MMR}).\\
In the analysis $\Tz$ is treated  as a free fit parameter. $\bar{T}_{\text{bl}}$ is kept at a fixed value which is obtained from MD simulations. From Fig.~\ref{Tepsilon} it can be seen that $\bar{T}_{\text{bl}} \simeq 0.5$ K for scenarios a and b. This indicates that the rf field used for secular excitation already induces heating of \Be\ while the field is still far away from the \Be\ resonance (at $\sim 300$~kHz). This effect is also seen in the experiment.

The nonlinear function $f_{\text{NL}}$ is used to map the relative \HD\ loss $\epsilon $ onto the spectroscopic signal $S$. However, for the correction of background signals (Sec.~\ref{bg}) we need to map $S$ to $\epsilon$, which requires the inverse non-linear function $f^{-1}_{\text{NL}}$. This inverse function is obtained numerically by use of $\textsc{Mathematica}$.

\section{Micromotion fit function}\label{MMfitfunction}
The time-dependent electric field of the trap, $\mathbf{E}_{\text{t}}$, can be expressed as ~\cite{Berkeland1998}
\begin{eqnarray}
 \mathbf{E}_{\text{t}}(x,y,z,t)& \cong &  -\frac{V_0}{R^2}(x \hat{x}-y\hat{y})\cos(\Omega t)\nonumber \\
 && -\frac{\kappa U_0}{Z_0^2}\times(2z\hat{z}-x\hat{x}-y\hat{y}) \label{eq:MMfit3},
\end{eqnarray}
where $R$ is half the distance between two diagonally opposing electrodes, $U_0$ is the endcap voltage, $Z_0$ stands for half the distance between the end caps, and $\kappa$ is a shielding factor. The Be$^+$ micromotion amplitude can be written as
\begin{equation}
\mathbf{x_{0}} = \frac{Q}{m_{\text{Be}} \Omega^2} \mathbf{E}_{\text{t}}(x,y,z,0). \label{eq:MMfit1}
\end{equation}
The measured micromotion amplitude can be written as
\begin{equation}
x_{0,k}=\frac{\mathbf{k}\cdot \mathbf{x_{0}}}{\|\mathbf{k}\|}\label{eq:MMfit2},%zoek correcte formule
\end{equation}
which is the projection of the 313 nm laser direction onto $\mathbf{x_{0}}$.
%$\mathbf{E}_{\text{t}}$ in Eq.~(\ref{eq:MMfit1}) is the $E$-field in the ion trap, which can be expressed as~\cite{Berkeland1998}
%\begin{eqnarray}
% \mathbf{E}_{\text{t}}(x,y,z,t)& \cong &  -\frac{V_0}{R^2}(x \hat{x}-y\hat{y})\cos(\Omega t)\nonumber \\
% && -\frac{\kappa U_0}{Z_0^2}\times(2z\hat{z}-x\hat{x}-y\hat{y}) \label{eq:MMfit3},
%\end{eqnarray}
%where $R$ is half the distance between two diagonally opposing electrodes, $U_0$ is the endcap voltage, $Z_0$ %stands for half the end cap to end cap distance, and $\kappa$ is a shielding factor.
From simulations of the rf trap circuitry with the simulation software \textsc{Spice}, we find a small possible phase difference $\phi_\text{ac}$ of 4~mrad in between the rf electrodes, which has a negligible effect on the ion micromotion and is ignored here. Using the program \textsc{SIMION}, we calculate the shielding factor $\kappa$ and the static electric fields $\mathbf{E_{\text{dc}}}$ as a function of the dc voltages applied to the trap electrodes. From the static electric fields, the radial ion displacement $r_d$ is obtained by balancing the ponderemotive force and static $E$-field in the radial direction,
\begin{equation}
m_{\text{Be}} \omega_r^2 r_d=-q E_{\text{dc}},
\end{equation}
where $\omega_{r}$ is the radial secular trap frequency. By inserting the $x$- and $y$-components of $r_d$ into Eq.~(\ref{eq:MMfit3}), the value of $E_{\text{t}}$ at the location of the ions is obtained.

A geometric imperfection of the trap could lead to an axial rf field, which can be written as (here we ignore the small modification of the radial rf field of the trap due to the same imperfection):
\begin{equation}
E_{\text{ax,HD}^+}(V_0,t)=\frac{1}{Q}\frac{V_0}{V_{\text{0,e}}} m_{\text{HD}} x_{\text{HD}}\Omega^2 \cos{\Omega t},
\end{equation}
where $x_{\text{HD}}$ is the HD$^+$ micromotion amplitude along the trap $z$-axis, $m_{\text{HD}}$ is the mass of \HD, and $V_{\text{0,e}}$ is the rf voltage used during the spectroscopic measurements, which is 270~V.

Now, we turn to the case of a linear string of Be$^+$ ions, which is the configuration used to determine the axial rf field amplitude. Adding $E_{\text{ax,HD}^+}$ to the $z$-component of $\mathbf{E}_{\text{t}}$ gives a new expression for $\mathbf{E}_{\text{t}}$ which is inserted into Eq.~(\ref{eq:MMfit1}).
We then obtain the following expression for  $\mathbf{x_{0}}$:
\begin{eqnarray}\label{x0vector}
\mathbf{x_{0}} & = & \Big( \frac{2 E_{\text{t},x}q R^2 V_0 Z_0^2}{q V_0^2 Z_0^2 -2 m_{\text{Be}} R^4 U_0 \kappa \Omega^2}, \nonumber \\
&& \frac{2 E_{\text{t},y} q R^2 V_0 Z_0^2}{q V_0^2 Z_0^2 -2m_{\text{Be}} R^4 U_0 \kappa \Omega^2},\nonumber \\
&& \frac{m_{\text{HD}}V_0 x_{\text{HD}}}{m_\text{Be} V_{0,\text{e}}} \Big),
\end{eqnarray}
which is subsequently inserted into Eq.~(\ref{eq:MMfit2}),
together with the wavevector, which is written as
\begin{equation}\label{kvector}
\mathbf{k} = \frac{2\pi}{\lambda}\left(\sin(\theta)\cos(\phi),\sin(\theta)\sin(\phi),\cos(\theta) \right).
\end{equation}
Here $\theta$ is the angle between $\mathbf{k}$ and the trap $z$-axis, and $\phi$ is the angle between $\mathbf{k}$ and the trap $y$-axis, which is very close to $\pi/4$ in our setup. The value of $\theta$ lies between $\pm 10$~mrad and is treated as a free fit parameter. We insert Eqs.~(\ref{kvector}) and (\ref{x0vector}) into Eq.~(\ref{eq:MMfit2})  and then expand the expression in powers of $\theta$. This gives us the following fit function:

\begin{eqnarray}\label{xHDthetaexp}
x_{0,k}(V_0)&=&\frac{m_{\text{HD}}}{m_{\text{Be}}}\frac{V_0 x_{\text{HD}}}{V_{\text{cal}}}\nonumber\\
&& -\frac{8(E_{\text{h,offs}}-E_{\text{v,offs}}+\delta E_{\text{h}}- \delta E_{\text{v}})Q^2 V_0}{m_{\text{Be}}R^2 \Omega^4 (2 a_{\text{M}}+q_{\text{M}}^2)}\theta \nonumber \\
&& + \mathcal{O}(\theta^2).
\end{eqnarray}
Here $E_\text{h,offs}$, $E_\text{v,offs}$ are the applied static electric fields in the horizontal and vertical directions, respectively, and $\delta E_{\text{h}}$, $\delta E_{\text{v}}$ are the unknown offset electric fields (due to \textit{e.g.} charging of electrodes). The Mathieu parameters $a_{\text{M}}$ and $q_{\text{M}}$ are given by
\begin{equation}\label{aq}
a_{\text{M}}=\frac{-4Q\kappa U_0}{m_{\text{Be}}Z_0^2 \Omega^2}, \quad q_{\text{M}}=\frac{2QV_0}{m_{\text{Be}}R^2 \Omega^2}.
\end{equation}
The displacement of the \Be\ string in the vertical direction can be accurately determined with images of the EMCCD camera, and therefore $\delta E_{h}$ can be zeroed (for example by minimizing the displacement of the Be$^+$ string while the radial confinement of the trap is modulated by varying the rf amplitude). However, the displacement in the horizontal direction (\textit{i.e.} perpendicular to the EMCCD image plane) is not accurately known and therefore we treat $\delta E_{\text{h}}$ as another free fit parameter. In summary, we use Eq.~(\ref{xHDthetaexp}) as a fit function with $x_{\text{HD}^+}$ , $\theta$ and $\delta E_{\text{h}}$ as free fit parameters while neglecting higher orders of $\theta$. The fitted curves and the result for $x_{0,k}$ are shown in Sec.~\ref{mm}.

The question arises what happens if the 782~nm laser propagates at a small angle with respect to the trap axis, while the \HD\ ions form a shell structure around the trap axis. In this case a small fraction of the radial micromotion is projected onto the wavevector. However, from Eqs.~(\ref{eq:MMfit1}-\ref{eq:MMfit3}) it follows that the sign of this additional micromotion alternates for each quadrant in the $(x,y)$ plane. As long as the radial micromotion component does not exceed the axial micromotion amplitude (which is the case here), the former averages out to zero given the radial symmetry of the \HD\ crystal.

\section{Stark shift calculations}\label{Starkshift}
Here we summarize the formulas that are used to calculate the ac Stark shift of a rovibrational transition $(v,L) \rightarrow (v',L')$ in the \HD\ molecule induced by a laser with intensity $I$ and polarization state $p$. A general expression for the second-order energy shift depending on the angle $\theta$ between the polarization direction and the quantization axis is:
\begin{equation}\label{eq:Stark1}
\begin{split}
\Delta E = & -\frac{1}{2}\frac{I}{c} [ \alpha^{(0)}_{vL}(\omega) \\
& + P_2(\cos \theta)\frac{3M^2-L(L+1)}{L(2L-1)}\alpha^{(2)}_{vL}(\omega) ],
\end{split}
\end{equation}
where $P_2(x)=\frac{1}{2}(3x^2-1)$ is a Legendre polynomial. This expression contains the scalar and tensor polarizabilities
\begin{eqnarray}\label{eq:Stark2}
\alpha^{(0)}_{vL}(\omega)&=&4\pi a^3_0 Q_s, \nonumber\\
\alpha^{(2)}_{vL}(\omega)&=&4\pi a^3_0 \sqrt{\frac{L(2L-1)}{(L+1)(2L+3)}}Q_t,
\end{eqnarray}
where $a_0$ is the Bohr radius and $Q_s$ and $Q_t$ stand for the two-photon scalar and tensor matrix elements:
\begin{eqnarray}\label{eq:Stark3}
Q_s&=& \frac{\langle vL\|Q^{(0)}\|v'L\rangle}{\sqrt{2L+1}}\nonumber\\
Q_t&=& \frac{\langle vL\|Q^{(2)}\|vL\rangle}{\sqrt{2L+1}}.
\end{eqnarray}
Here $Q^{(0)}$ and $Q^{(2)}$ are the irreducible scalar and tensor components that belong to the two-photon operator (in atomic units):
\begin{equation}\label{eq:Stark4}
Q_{pp}(E)=\mathbf{d} \cdot \mathbf{\epsilon}_{\mathbf{p}}\frac{1}{H-E}\mathbf{d} \cdot \mathbf{\epsilon}_{\mathbf{p}},
\end{equation}
with Hamiltonian $H$, dipole moment operator, $\mathbf{d}$  and polarization vector  $\mathbf{\epsilon}_{\mathbf{p}}$.
The matrix elements $Q_s$ and $Q_t$ were calculated numerically using the three-body variational wave functions described in~\cite{Karr2014}.

Since the hyperfine structure is partially resolved in this spectrum, we also have to consider the contribution of the Stark shifts to off-resonant coupling to hyperfine levels in $v=0$ and $v=8$ by the 782~nm laser during spectroscopy. Here, the situation is more complicated as the 782~nm laser also non-resonantly couples $v=8$ states to continuum states above the dissociation limit of the 1s$\sigma$ electronic ground state. The 782-nm contribution to the Stark shift was calculated at the hyperfine level, and will be published elsewhere. The corresponding shifts turn out to be negligible for our experiment, contributing only at the level of a few Hertz.

%\nocite{*}
\bibliographystyle{my_ieeetr}
\bibliography{bibPRAJB1}% Produces the bibliography via BibTeX.

\end{document}